\documentclass{aa}  

\usepackage{amsmath}
\usepackage{amssymb}
\usepackage{txfonts}
\usepackage{graphicx}
\usepackage{gensymb}
\usepackage{textcomp}
\usepackage[colorlinks=true, linkcolor=blue, citecolor=blue, urlcolor=blue]{hyperref}
\usepackage{natbib}
\usepackage{tabularx}

\bibpunct{(}{)}{;}{a}{}{,} 
\DeclareMathAlphabet{\mathpzc}{OT1}{pzc}{m}{it}

\begin{document} 

\title{High-resolution, 3D radiative transfer modelling}
\subtitle{II. The early-type spiral galaxy M~81}

\author{
Sam~Verstocken\inst{\ref{UGent}}
\and
Angelos~Nersesian\inst{\ref{UGent}}\fnmsep\inst{\ref{NOA}}\fnmsep\inst{\ref{UAth}}
\and
Maarten~Baes\inst{\ref{UGent}}
\and
S\'ebastien~Viaene\inst{\ref{UGent}}\fnmsep\inst{\ref{Herts}}
\and
Simone~Bianchi\inst{\ref{Arcetri}}
\and
Viviana~Casasola\inst{\ref{Bologna}}\fnmsep\inst{\ref{Arcetri}}
\and
Christopher~J.~R.~Clark\inst{\ref{STScI}}
\and
Jonathan~I.~Davies\inst{\ref{Cardiff}}
\and
Ilse~De~Looze\inst{\ref{UGent}}\fnmsep\inst{\ref{UCL}}          
\and
Pieter~De~Vis\inst{\ref{Cardiff}}
\and
Wouter~Dobbels\inst{\ref{UGent}}
\and
Fr\'ed\'eric~Galliano\inst{\ref{CEA}}
\and
Anthony~P.~Jones\inst{\ref{IAS}}
\and
Suzanne~C.~Madden\inst{\ref{CEA}}
\and
Aleksandr~V.~Mosenkov\inst{\ref{RAS}}\fnmsep\inst{\ref{StPSU}}
\and
Ana~Tr\v{c}ka\inst{\ref{UGent}}
\and
Emmanuel~M.~Xilouris\inst{\ref{NOA}}
}
		  
\institute{
Sterrenkundig Observatorium, Universiteit Gent, Krijgslaan 281 S9, B-9000 Gent, Belgium
\label{UGent}
\and
National Observatory of Athens, Institute for Astronomy, Astrophysics, Space Applications and Remote Sensing, Ioannou Metaxa and Vasileos Pavlou GR-15236, Athens, Greece
\label{NOA}
\and
Department of Astrophysics, Astronomy \& Mechanics, Faculty of Physics, University of Athens, Panepistimiopolis, GR15784, Zografos, Athens, Greece
\label{UAth}
\and
Centre for Astrophysics Research, University of Hertfordshire, College Lane, Hatfield AL10 9AB, UK
\label{Herts}
\and
INAF -- Osservatorio Astrofisico di Arcetri, Largo E. Fermi 5, 50125, Firenze, Italy
\label{Arcetri}
\and
INAF -- Istituto di Radioastronomia, Via P. Gobetti 101, 4019, Bologna, Italy
\label{Bologna}
\and
Space Telescope Science Institute, 3700 San Martin Drive, Baltimore, Maryland, 21218, USA
\label{STScI}
\and
School of Physics and Astronomy, Cardiff University, Queen's Buildings, The Parade, Cardiff, CF24 3AA, UK
\label{Cardiff}
\and
Department of Physics and Astronomy, University College London, Gower Street, London WC1E 6BT, UK
\label{UCL}
\and
AIM, CEA, CNRS, Universit\'e Paris-Saclay, Universit\'e Paris Diderot, Sorbonne Paris Cit\'e, 91191, Gif-sur-Yvette, France
\label{CEA}
\and
Institut d'Astrophysique Spatiale, CNRS, Universit\'e Paris-Sud, Universit\'e Paris-Saclay, B\^at.\ 121, 91405, Orsay Cedex, France
\label{IAS}
\and
Central Astronomical Observatory of RAS, Pulkovskoye Chaussee 65/1, 196140, St. Petersburg, Russia
\label{RAS}
\and
St. Petersburg State University, Universitetskij Pr. 28, 198504, St. Petersburg, Stary Peterhof, Russia
\label{StPSU}
}

\date{Received 25 April 2019; accepted 7 April 2020}
   
\offprints{M. Baes, \email{maarten.baes@ugent.be}}

\abstract
{Interstellar dust absorbs stellar light very efficiently and thus shapes the energetic output of galaxies. Studying the impact of different stellar populations on the dust heating remains hard because it requires decoupling the relative geometry of stars and dust, and involves complex processes as scattering and non-local dust heating.}
{We aim to constrain the relative distribution of dust and stellar populations in the spiral galaxy M~81 and create a realistic model of the radiation field that describes the observations. Investigating the dust-starlight interaction on local scales, we want to quantify the contribution of young and old stellar populations to the dust heating. We aim to standardise the setup and model selection of such inverse radiative transfer simulations so this can be used for comparable modelling of other nearby galaxies.}
{We present a semi-automated radiative transfer modelling pipeline that implements the necessary steps such as the geometric model construction and the normalisation of the components through an optimisation routine. We use the Monte Carlo radiative transfer code SKIRT to calculate a self-consistent, panchromatic model of the interstellar radiation field. By looking at different stellar populations independently, we can quantify to what extent different stellar age populations contribute to the dust heating. Our method takes into account the effects of non-local heating.}
{We obtain a realistic 3D radiative transfer model of the face-on galaxy M~81. We find that only 50.2\% of the dust heating can be attributed to young stellar populations ($\lesssim 100\,\text{Myr}$). We confirm a tight correlation between the specific star formation rate and the heating fraction by young stellar populations, both in sky projection and in 3D, also found for radiative transfer models of M~31 and M~51.}
{We conclude that the old stellar populations can be a major contributor to the heating of dust. In M~81, old stellar populations are the dominant heating agent in the central regions and they contribute to half of the absorbed radiation. Regions of higher star formation do not correspond to the highest dust temperatures. On the contrary, it is the dominant bulge which is most efficient in heating the dust. The approach we present here can immediately be applied to other galaxies. It does contain a number of caveats which are discussed in detail.}
   

\maketitle

\section{Introduction}

Interstellar dust reprocesses optical and UV photons into infrared (IR) and submillimetre (submm) radiation. The line-of-sight extinction and reddening effects of dust in optical and UV wavebands in the Milky Way and other galaxies has provided us with crucial insights regarding the nature of absorption and scattering processes, as well as the dust grain composition \citep{Draine2007, Calzetti2000}. On the other hand, observations at mid-infrared (MIR), far-infrared (FIR) and submm wavelengths have provided constraints on dust density, temperature, emissivity and the correlation between the presence of dust and star formation. Since the advent of missions such as {\textit{Spitzer}} \citep{Werner2004} and {\textit{Herschel}} \citep{Herschel}, it has become possible to study interstellar dust in hundreds of nearby galaxies on resolved scales. This has allowed quantifying the properties for different regions of such galaxies in different environments. For a recent review on the interstellar dust properties in nearby galaxies, see \citet{2018ARA&A..56..673G}.

Studies based on these data have shown that the diffuse, old stellar population in galaxies can play a significant role in the heating of interstellar dust \citep{Hinz2004, Bendo2010, Bendo2012, 2015MNRAS.448..135B}, whereas star formation can be attributed to most of the dust emission at shorter wavelengths, e.g., at 24 $\mu$m \citep{Calzetti2005, Calzetti2007, Prescott2007, Kennicutt2007, 2009ApJ...703.1672K, Zhu2008}. \citet{Lu2014} find that, in the case of M~81, even at 8~$\mu$m and 24~$\mu$m the old stars are responsible for 67\% and 48\% of the dust emission in the respective bands. 

\citet{2019A&A...624A..80N} have investigated the dust-starlight interplay in 814 DustPedia\footnote{DustPedia \citep{Davies2017, Clark2018} is multi-wavelength survey of galaxies in the Local Universe. The DustPedia sample consists of 875 nearby galaxies (recessional velocities of $< 3000$~km~s$^-1$) with {\textit{Herschel}} detection and angular sizes of $D_{25} > 1\arcmin$.} galaxies across all Hubble stages using the spectral energy distribution (SED) fitting code CIGALE \citep{2019A&A...622A.103B}. They found that the relative contribution of young stellar populations ($\leqslant 200$~Myr) to the dust heating is monotonically increasing with Hubble Stage, providing --on average-- only 10\% of the absorbed energy in early-type galaxies up to 60\% in the later type galaxies. 

Using the global SED of a galaxy, however, only permits an average picture of the dust heating mechanisms within the galaxy. Therefore, the correlation between star formation and IR emission is stronger in certain regions of a galaxy, underestimated by the global fraction of heating by young stars as indicated by the global SED model. Moreover, global SED models assume dust absorbs light from either stellar population independent of their relative spatial distribution. Differences in heating fraction between galaxies, therefore, are the direct result of the relative amount of young and old stars required to explain the UV-NIR part of the galaxy spectrum, and how these intrinsic stellar spectra vary as a function of position over the galaxies. 

To take into account the geometry of stellar and dust components, SED fitting tools can also be used on resolved scales, by fitting the spectrum on a pixel-by-pixel basis. This requires a panchromatic dataset defined on a uniform coordinate grid. These methods have proven useful for studying local variations in star formation, dust density, dust characteristics, etc. \citep{2011A&A...536A..88G, Gordon2014, Viaene2014, 2019MNRAS.486..743D}. However, SED fitting still has its limitations in this context, as each pixel is treated independently. This ignores the fact that, depending on the inclination, the line-of-sight probes different regions in the galaxy, potentially bearing different stellar and dust properties. More importantly, enforcing a local energy balance does not take into account the propagation of radiation over lengths greater than the physical scale of the pixels. This non-local character of dust heating might be important \citep{Smith2015, Boquien2015}. 3D radiative transfer analyses have demonstrated the non-locality of dust heating in the Sombrero galaxy \cite{DeLooze2012Sombrero} and the Andromeda galaxy \citep{V17}. In both galaxies, the radiation from the old stellar populations in the bulge has proven to propagate far outwards and contribute to the heating out to large distances.

To include these non-local dust heating effects, a self-consistent 3D model of the galaxy is required that fully incorporates the effects of scattering, absorption, and dust emission. Bearing the cumulative effect of these physical processes, the photons must be propagated back into the line-of-sight to make a proper comparison to observed galaxies possible. Thus, a 3D panchromatic radiative transfer (RT) model is required to obtain a realistic description of the stellar and dust distribution (see \citealt{SteinackerBaes2013} for an overview of RT codes and description of the techniques). 

The procedure described above, where the 3D stellar and dust compositions are simultaneously constrained by fitting the emergent radiation field to observed data of a galaxy, is referred to as inverse 3D radiative transfer or 3D radiative transfer modelling. Up to now, most attempts at RT modelling entire galaxies focused on edge-on spiral galaxies and typically adopted axial symmetry \citep[e.g.,][]{Xilouris1999, Bianchi2008, Baes2010, Popescu2011, 2012ApJ...746...70S, DeLooze2012ngc4565, DeLooze2012Sombrero, Mosenkov2016, Mosenkov2018}. In their RT study of M~51, \citet{DL14} introduced novel techniques for the modelling of well-resolved face-on galaxies, and effectively marked the onset of full 3D radiative transfer modelling. The main advantage of using face-on galaxies is that the filamentary and spiral features observed in the disc can be directly used for the geometric model construction, whereas for edge-on galaxies only 2D axially symmetric, smooth discs can be assumed. The use of these heterogeneously mixed geometries, and the inclusion of multiple UV components allow for a much more realistic analysis of the interaction between stellar radiation and the interstellar matter, which reduces the dust energy balance problem \citep{Saftly2015}. 

This work is a continuation and extension of the analysis of \citet{DL14}, and also aims to set the strategy for a systematic radiative transfer modelling effort of a larger sample of nearby galaxies \citep[e.g.,][]{Nersesian2020, Viaene2020}. Within the scope of DustPedia \citep{Davies2017}, with consistency and reproducibility in mind, we have bundled the modelling tools into a software package that is specifically aimed at convenient future use, and designed to be able to tackle a larger sample of face-on galaxies. In this work, we will describe this modelling approach and apply it to the early-type spiral galaxy M~81 (NGC~3031, Sab, Hubble stage = 2.4, $D\approx 3.7$~Mpc, $i \approx 59$~deg). M~81 is an interacting galaxy \citep[e.g.,][]{Casasola2004}, belonging to the M~81 group consisting of 34 members. The gas content of M~81 has been extensively studied \citep[e.g.,][]{Combes1977, Sakamoto2001, Casasola2007, Heiner2008, SanchezGallego2011} because of the strong density wave attributed to the tidal interaction with the companions of the M~81 group, especially M~82 and NGC~3077 \citep{Kaufman1989}.

This paper is structured as follows. In Sect.~\ref{sec:data}, we describe how the multi-wavelength image data for M~81 were gathered and prepared for the model construction and verification. In Sect.~\ref{sec:approach}, we describe in detail our updated approach for the setup of a 3D radiative transfer model of face-on galaxies and show the specific properties of the model of M~81. The results of the M~81 modelling are described in Sect.~\ref{sec:results}, including the verification of the model based on its SED and an investigation in the dust heating mechanisms. In Sect.~{\ref{sec:discussion}} we discuss the implications of our modelling results, and we discuss the limitations and caveats in our modelling approach. The conclusions are listed in Sect.~\ref{sec:conclusions}. 

\section{Data collection and preparation}
\label{sec:data}

For M~81, imaging data are available from all of the 7 telescopes for which data are gathered as part of DustPedia ({\textit{GALEX}}, SDSS, 2MASS, {\textit{WISE}}, {\textit{Spitzer}}, {\textit{Herschel}}, and {\textit{Planck}}). This gives 35 images, which are automatically retrieved from the DustPedia archive\footnote{\url{http://dustpedia.astro.noa.gr}} by the modelling pipeline. For an overview of the pixel scales, FWHMs and calibration uncertainties of the DustPedia images, see \citet{Clark2018}. In addition to the images that are part of the sample, a continuum-subtracted H$\alpha$ image is retrieved from NED\footnote{\url{https://ned.ipac.caltech.edu}}, from the study of \citet{Hoopes2001}.

We removed stars from {\textit{GALEX}}, SDSS, and {\textit{Spitzer}} IRAC images with an automatic procedure, developed as part of the modelling pipeline and also described briefly by \citet{Clark2018}. The procedure, in broad terms, consists of the following steps. First, the 2MASS All-Sky Catalog of Point Sources \citep{Cutri2003} is queried for the coordinate range of the particular image. The catalog data is accessed through the Vizier interface of the \textsc{Astroquery} package \citep{Astroquery}, an \textsc{Astropy} affiliated package \citep{Astropy2018}. For each position in the catalog, a local peak above a 3-sigma significance level is looked for. Non-detections result in the catalog position being ignored. The neighbouring pixels of each remaining point source position are then subtracted by the local background and fitted to a 2D Gaussian profile. If fitting succeeded, a segmentation step is performed to detect saturation bleed, diffraction patterns and ghosts typically found around the brightest stars. The flagged pixels are then interpolated over by replacing them by a weighted average of the neighbouring pixels that are not flagged, complemented with random noise mimicking the variation in these pixels. Manual inspection is performed afterwards and the procedure is repeated with mistakenly identified point sources ignored or additional regions flagged for interpolation.

\begin{figure*}[t!]
\centering
\includegraphics[width=\textwidth]{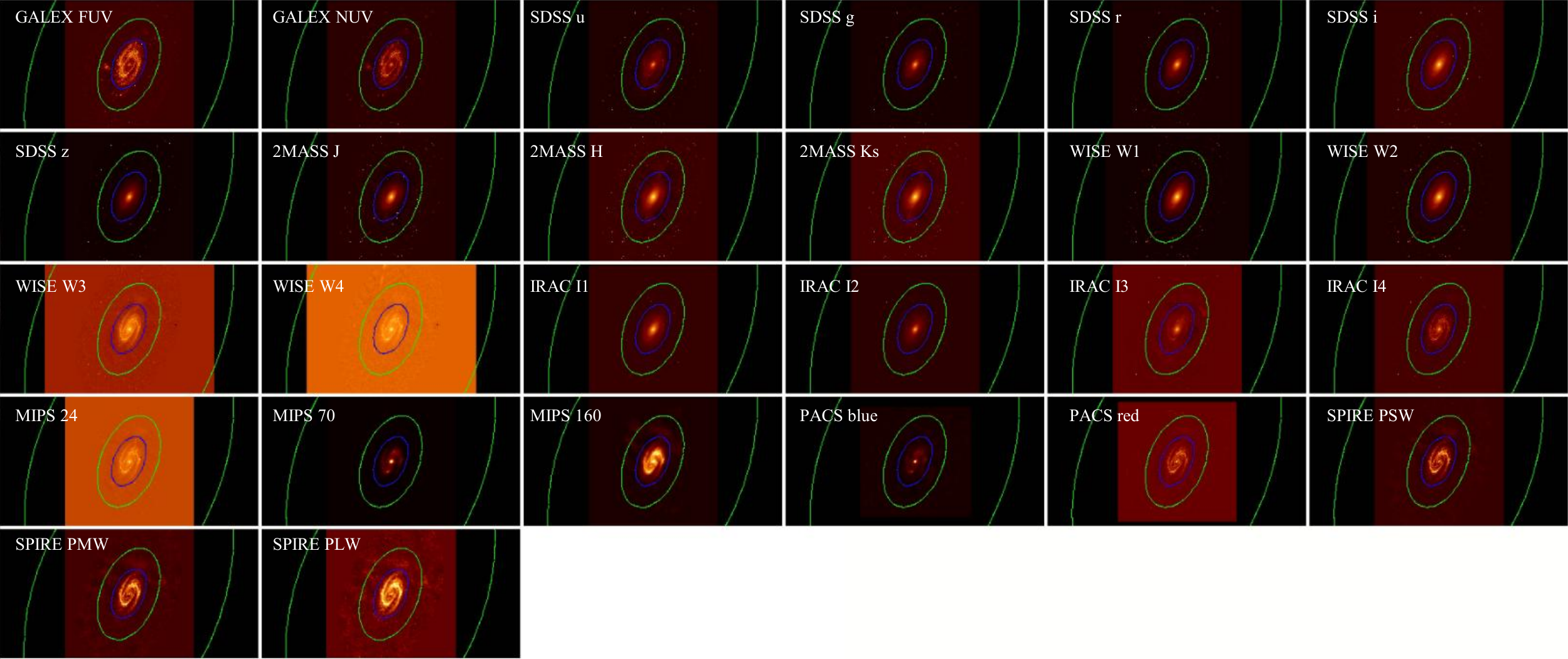}
\caption{Photometry thumbnail image grid for M~81. The blue ellipse in each image is the master ellipse used for the determination of the integrated flux. The green ellipses mark the annulus used for the determination of the background.}
\label{fig:aperturephotometry_M81}
\end{figure*}

All images are corrected for background emission and systematic offsets by estimating the large-scale variation of pixel values around the galaxy. We use the mesh-based background estimation tools from \textsc{Photutils} \citep{Photutils}. In this method, the image is divided up into a grid of rectangular regions where the local background is estimated, after which this grid is upsampled to the original pixel resolution. We have chosen square boxes with a side of six times the FWHM of the image. All pixels identified as belonging to the galaxy (as an elliptical region) are masked, as well as all pixels flagged during the source extraction as being either a foreground star or a background galaxy. To estimate the sky and its standard deviation behind the galaxy, the maps resulting from the \textsc{photutils} upsampling method are interpolated within the central galaxy ellipse.

Images affected by Milky Way attenuation must be corrected by a factor depending on the filter response curve and a galactic extinction law \citep{Cardelli1989}. We calculate the attenuation in all wavebands $<10 \, \mu \text{m}$ by convolving a $R_\lambda$ curve with $R_V=3.1$ with the corresponding filter transmission curves obtained from the SVO Filter Profile Service\footnote{\url{http://svo2.cab.inta-csic.es/svo/theory/fps/}}, and using the \textsc{IrsaDust} tool of \textsc{Astroquery} (querying the IRSA Dust Extinction Service\footnote{\url{https://irsa.ipac.caltech.edu/applications/DUST/}}) to obtain the V-band attenuation $A_V$ for the galaxy position \citep{2011ApJ...737..103S}. This step is also automated within the modelling pipeline.

For {\textit{Spitzer}} and {\textit{Herschel}} images, error maps are available from the DustPedia archive, derived from their respective data reduction pipelines. To assert uniformity across our multi-wavelength dataset, however, we opt to generate the error maps independently with our modelling pipeline. In general, the error map for a certain band is a combination of three contributions: the calibration uncertainty, pixel-to-pixel noise estimated from the deviation from the smooth background, and Poisson noise. The pixel-to-pixel noise is the deviation of the pixel values from the assumed smooth background, which is produced during the sky subtraction step. The Poisson noise is taken into account for photon-counting instruments where the number of detected photons in an individual pixel can be low. We create Poisson error maps for the {\textit{GALEX}} and SDSS images. Creating these maps is not trivial, as the DustPedia {\textit{GALEX}} and SDSS images are the result of a mosaicking of individual observations. In the case of {\textit{GALEX}}, we have to take into account a variable exposure time for each observation. For SDSS, the observations (fields) have to be recalibrated from nanomaggy units into photon counts, taking into account the 2D polynomial sky that has been subtracted by the SDSS imaging pipeline and a different calibration factor per image column. The pixel noise and Poisson noise (where applicable) are added in quadrature with the calibration error map for each band. The relative calibration errors for each band are listed in Table~1 of \citet{Clark2018}.

Global photometry is applied to each clipped image by summing all pixels within a fixed elliptical aperture with semi-major and semi-minor axes of 727\farcs7 and 421\farcs2, respectively. An overview of the different images with the apertures for the integrated flux determination and the annuli for the determination of the background is provided in Fig.~{\ref{fig:aperturephotometry_M81}}. We have selected this aperture to exclude extended emission that is present in some images and the satellite dwarf galaxy Homberg IX that is most prominent in the UV bands. Corresponding uncertainties are also calculated by adding the calibration uncertainty and the mean variation from the background in quadrature. The resulting flux densities, listed in the table in the Appendix, differ only slightly from the DustPedia photometry (on average, our fluxes are about 2\% lower). The detection of M~81 in the {\textit{Planck}} images is extremely faint compared to the Galactic cirrus or even indistinguishable from it, which is why our HFI~100~GHz, HFI~143~GHz and LFI fluxes are upper limits. The {\textit{Planck}} images and fluxes are not used to construct the RT model and thus serve only for visual comparisons.

\section{Modelling approach}
\label{sec:approach}

\subsection{General outline}

\begin{figure}[t!]
\centering
\includegraphics[width=\columnwidth]{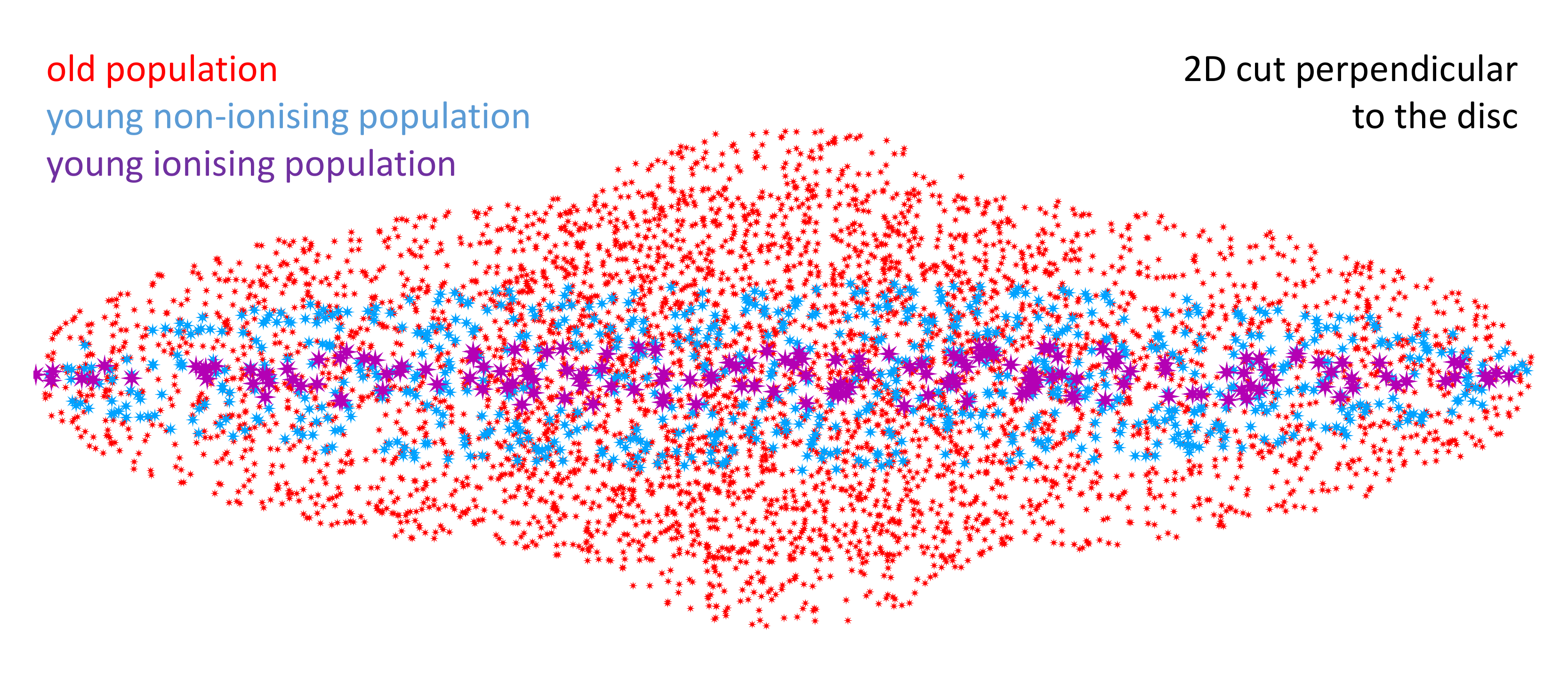}
\caption{The standard geometric model of spiral galaxies adopted in our models. The old stellar population consists of a disc and bulge component. The young stellar population is assumed to be restricted to the disc and consists of a non-ionising subpopulation with an age around 100~Myr, and an ionising subpopulation of stars with an age of 10~Myr. The young non-ionising population resides in a thinner disc compared to the old stars, and the young ionising population is even more compacted to the mid-plane of the galaxy. The dust distribution exhibits the same vertical extent as the young non-ionising stellar population.}
\label{fig:geometry}
\end{figure}

In this section, we present a general overview of our modelling strategy. The interested reader will find a detailed description in sections~\ref{ssec:components} and \ref{ssec:skirt}. Other readers may continue to our optimisation strategy (Sect.~\ref{ssec:optimisation}).

The general modelling approach that we apply is the following. We construct a 3D model for the galaxy, consisting of several stellar components and a dust component. More specifically, we consider four different stellar components: a central bulge of old stars, an old stellar disc, a young non-ionising stellar disc, and a disc with an even younger stellar population (the young ionising disc). This standard model with three distinct stellar age categories is illustrated in Fig.~\ref{fig:geometry}. 

The geometrical distribution of each of the components is based on observed images of the galaxy at different wavelengths, using a two-step procedure. The first step consists of combining different images to physical maps that characterise, for example, the distribution of dust or old stellar populations on the sky. The second step consists of de-projecting these 2D maps on the sky to a 3D distribution. Apart from the geometrical distribution, each stellar component is assigned an intrinsic SED, and a total luminosity that is either fixed or a free parameter in the model. Similarly, the optical properties of the dust component are specified, and the total dust mass is a free parameter in the model. 

For any choice of the three free parameters in the galaxy model, we can set up a 3D dust radiative transfer simulation that fully takes into account the emission by the different stellar populations, and the absorption, scattering and thermal emission by the dust. These simulations are performed with SKIRT, a general-purpose dust radiative transfer code based on the Monte Carlo technique \citep{Baes2011, CampsBaes2015} that is publicly available \footnote{\url{http://www.skirt.ugent.be}}. The challenging 3D galaxy simulations we run here are possible thanks to the large suite of input models \citep{BaesCamps2015}, the advanced grid structures \citep{Saftly2013, Saftly2014}, the optimisation techniques \citep{SteinackerBaes2013, Baes2016}, and the hybrid parallelisation strategies \citep{Verstocken2017} implemented in SKIRT.

The result of each such simulation is a well-sampled 3D ($x,y,\lambda$) cube from which we extract the SED and a set of broadband images of the galaxy. These images, from UV to submm wavelengths, can directly be compared to the observed images. The best fitting model is determined through a $\chi^2$ optimisation procedure: using a two-step grid-based fitting approach, we determine the free parameters of the model that reproduces the observed SED best. As soon as the best fitting model is determined, a wealth of possible information can be extracted. For example, additional images of the galaxy at arbitrary viewing points and arbitrary wavelengths can be calculated, and the intrinsic properties of the interaction between starlight and dust can be studied in 3D. 

In general, our modelling procedure largely follows the same approach as \citet{DL14} and \citet{V17}, but with some improvements and important differences. One major distinction is that we construct all the components in the model at their ``native'' resolution, in the sense that we no longer re-bin and convolve all input images to the same resolution before we use them to construct our stellar and dust components. Instead, we ensure that we match the resolution of each particular set of images that is used for a specific map, and as such allow for the highest possible resolution for each particular component. Another difference is the computation of a pixel-by-pixel goodness-of-fit metric across all bands instead of solely using fluxes from the integrated SED.

Overall, we have taken special care to automate the entire procedure as much as possible. The different steps in the procedure, discussed in the following subsections, are implemented in Python and are publicly available as part of the Python Toolkit for SKIRT (PTS)\footnote{\url{http://www.skirt.ugent.be/pts}}. This automation and open-science approach has the obvious advantages that our results can immediately be reproduced or extended by other teams, and that the same approach can easily be applied to other galaxies. In the subsections below, we provide more details on the different modelling steps, complemented by aspects that are specific to the model of M~81.

\subsection{The different components in the model}
\label{ssec:components}

\subsubsection{The stellar bulge}

The stars and dust in a spiral galaxy are distributed in a flattened disc, with the exception of a central bulge that consists primarily of old stellar populations \citep[e.g.,][]{Alton1998, Bianchi2007, MunozMateos2009, Hunt2015, Casasola2017}. The first step in the modelling procedure is therefore to set up a suitable model for the bulge. The old stellar population can be best seen in NIR wavebands, where the contribution from recently formed stars is usually limited, the contamination from dust emission is negligible, and internal extinction by the dust is also modest. 

The {\textit{Spitzer}} Survey of Stellar Structure in Galaxies \citep[S$^4$G:][]{Sheth2010, Salo2015} has provided a database of decompositions in {\textit{Spitzer}} IRAC~3.6 and 4.5~$\mu$m bands for 2352 nearby galaxies. The bulges of the galaxies are usually modelled as flattened S\'ersic models, and the decomposition parameters can be retrieved from an ASCII table provided on the S$^4$G pipeline webpage\footnote{\url{http://www.oulu.fi/astronomy/S4G_PIPELINE4/MAIN/}}. 

The 2D S\'ersic surface brightness distribution is de-projected to a 3D intrinsic distribution under the assumption that the bulge has an oblate spheroidal distribution, and that the inclination of the bulge is the same as the inclination of the old stellar disc (we assume that all components have the same inclination). The SKIRT code is equipped with a dedicated routine ({\texttt{SersicGeometry}}) that constructs a 3D spheroidal bulge model, based on the characteristics of the S\'ersic surface brightness distribution and the inclination. The spatial density and cumulative mass distribution can be expressed analytically using special functions \citep{BaesGentile2011, BaesvanHese2011}.

The setup of a stellar component is not complete until the intrinsic spectrum and the normalisation are defined. For the old stellar population in the bulge, we adopt a \citet{BruzualCharlot2003} spectral energy distribution with a fixed metallicity and an age of 8~Gyr. These numbers are similar to the values used by \citet{V17} for the bulge of M~31, but they can obviously be changed if this information is available for the particular galaxy to be modelled. The old stellar age of 8~Gyr is a middle ground between 5-13~Gyr, the common range for old stellar populations. In fact, the actual SEDs don't change too much within this age range. \citet{Kong2000} find a super-solar metallicity of 0.03 for M~81 that is rather uniform across the galaxy, which is therefore the value we adopt for the stellar components of our M~81 model.

The total bulge luminosity is fixed by the total IRAC~3.6~$\mu$m flux density from the bulge as obtained from the best fitting S$^4$G model, and this luminosity is assumed to be unaffected by internal dust attenuation and thus can be directly used as the intrinsic normalisation of the bulge component.

\begin{figure*}[t!]
\centering
\includegraphics[width=\textwidth]{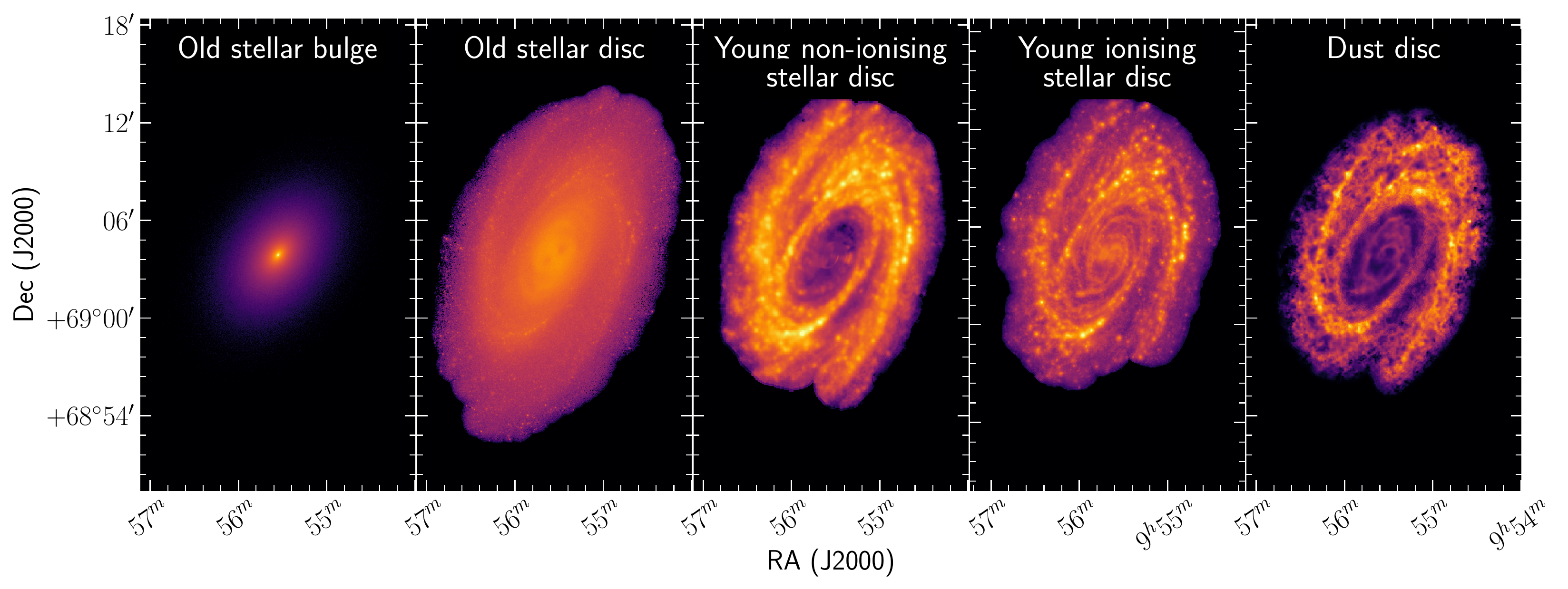}
\caption{A 2D view of the different model components, including a bulge and disc component for the old stellar population, discs of young non-ionising and ionising stellar populations, and a disc of the dust distribution. The bulge image has been rendered with SKIRT by using the 3D bulge description and the appropriate instrument setup. The disc component maps have been produced with recipes directly using the observed multi-wavelength images, as detailed in the text. The different maps have different native resolutions based on the respective observations that were used to create these maps. These resolutions are 1\farcs9 for the old stellar bulge and disc, 11\farcs2 for the young non-ionising stellar disc, 6\farcs4 for the young ionising stellar disc, and 11\farcs2 for the dust disc (see Tab.~\ref{tab:maps}). The removal of unphysical pixels due to different signal-to-noise in each of the bands leads to a varying contour for the different components.}
\label{fig:maps}
\end{figure*}

For the specific case of M~81, the S$^4$G team have found a best-fitting S\'ersic model for the bulge with a S\'ersic index $n=3.56$, an effective radius of 1.38~kpc, an axial ratio (flattening) of 0.654, and a position angle of 55~degrees \citep{Salo2015}. The IRAC~3.6~$\mu$m flux density of the bulge is 5.31~Jy, corresponding to a luminosity of 
\begin{equation}
L_{\text{bulge},\,3.6} = 2.11\times10^{35}~{\text{W}}~\mu{\text{m}}^{-1}
\end{equation}
We use SKIRT to render an image of the bulge using the \texttt{SersicGeometry} and arbitrary normalisation to obtain a 2D image of the bulge component on the plane of the sky. This image is shown in the first panel of Fig.~\ref{fig:maps}.

\subsubsection{The old stellar disc}

A map of the old stellar population in the disc of the galaxy can be obtained by subtracting the bulge (as resulting from the decomposition) from the total observed image in the IRAC~3.6~$\mu$m band. With the technique introduced by \citet{DL14} and implemented in SKIRT's {\texttt{ReadFitsGeometry}}, we can directly load this 2D map into SKIRT which converts it into a 3D disc component. De-projections such as these are generally degenerate and require some assumptions to generate a unique solution. We resolve the degeneracy by imposing that the old stellar disc has a perfectly exponential distribution in the vertical direction, with a fixed scale height (actually, we make the same assumption for all the disc components, with a separate scale height for each component) \citep{Wainscoat1989, Wainscoat1990, vanderKruit2011}. The method works by mapping each pixel in the image to a position in the mid-plane of the 3D model, implying a rotation over the position angle and essentially a stretch of the image along the direction of the minor axis with a factor of $\cos i$. The direct mapping from plane of the sky to plane of the galaxy makes the implicit assumption of an infinitely thin disc, but is also a reasonable approximation for low inclination angles. \citet{V17} have even achieved realistic de-projection results for M~31 with an inclination angle of 77.5$^\circ$. We base our estimated inclination angle for M~81 on the apparent flattening of the disc found by the S$^4$G decomposition, using the Hubble formula \citep{Hubble1926}:

\begin{equation}
\label{eq:hubble}
\sin^2 i = \frac{1-q^2}{1-q_0^2}
\end{equation}

An important parameter is the exponential scale height of the disc. Various studies of edge-on disc galaxies have found quite strong correlations between the scale height of the stellar component and other properties, including the stellar disc scale length, total luminosity, and rotational velocity \citep[e.g.,][]{deGrijs1998, Kregel2002, Bianchi2007, Courteau2007}. We use the results obtained by \citet{DeGeyter2014} based on detailed radiative transfer modelling of 12 edge-on spiral galaxies. They found a mean intrinsic flattening, i.e. ratio of scale length to scale height, of 8.26. This value is almost exactly the same as the value obtained by \citet{Kregel2002} based on independent surface brightness model fitting of 34 edge-on spiral galaxies. Our strategy is thus to take the scale length as obtained from the S$^4$G bulge-disc decomposition, and use that to compute a value for the scale height of the old stellar disc. 

Finally, we generally assume the same stellar population (an SSP of 8~Gyr and a fixed metallicity) as for the old population of the bulge, and we fix the total luminosity based on the fitted disc luminosity from the S$^4$G exponential disc fit. 

For the specific case of M~81, the old stellar disc map is shown in Fig.~\ref{fig:maps}. From S$^4$G, we find a scale height of 332~pc and a total IRAC~3.6~$\mu$m luminosity slightly higher than that of the bulge:

\begin{equation}
L_{\text{disc},\,3.6} = 2.20\times10^{35}~{\text{W}}~\mu{\text{m}}^{-1}
\end{equation}

Using the intrinsic flattening $q_0=1/8.26$ and an apparent axis ratio of 0.543, formula \ref{eq:hubble} yields an inclination angle $i=57.8^\circ$.

\subsubsection{The young non-ionising stellar disc}

Apart from old stellar populations, galaxies also contain recently formed stars. These stellar populations constitute a minor fraction of total stellar mass, but as they dominate the emission at blue and UV wavelengths, they play a crucial role in the heating of the interstellar dust and thus in our RT models. Similar to \citet{DL14} and \citet{V17}, we separated the young stellar population into two subpopulations. The main subpopulation, which we denote as the young non-ionising population, corresponds to stars that have formed about 100 Myr ago and that have already left their birth clouds. The second subpopulation, which we denote as the young ionising population, corresponds to stars formed within the last 10~Myr, which are therefore assumed to be still embedded in their birth clouds, heating it by their hard ionising radiation. Splitting the young population into these two disc subcomponents allows for dust in clumpy regions, (perhaps) not picked up in attenuation from the diffuse stars, but still contributing in emission due to the contained star formation activity. In fact, this clumpy dust will be incorporated as a sub-grid feature, as discussed in \ref{sssec:ionising}. Thus, this method guards against energy balance problems, without necessarily imposing a fixed fraction of young stars to be ionising or introducing artificial clumping in the dust geometry. 

The population of young stars that are no longer obscured by their birth clouds of gas and dust can be expected to dominate the emission in the FUV band, and to a lesser extent the NUV band, although there may be a significant contribution from the older stellar population, certainly in the inner regions. Additionally, FUV radiation is very efficiently absorbed and scattered (attenuated) by dust and therefore the intrinsic luminosity of the young non-ionising stellar population is not directly related to the observed FUV flux density. 

The first step in the construction of an intrinsic 3D model for the distribution of the young non-ionising stellar population consists of making a map of the intrinsic FUV emission. To create this map, we combine the observed FUV surface brightness map with an estimate of the FUV attenuation map. This attenuation map is set up using the prescriptions of \citet{Cortese2008}, who have performed a regression of the FUV attenuation as a polynomial function of the logarithm of the total infrared (TIR) to FUV ratio, on spatially resolved scales. The coefficients of the polynomial are determined for different values of the specific star formation rate (sSFR). Following the prescriptions described in \citet{Galametz2013}, we determine a TIR emission map based on the MIPS 24 $\mu$m, the PACS~70~$\mu$m, and the PACS~160~$\mu$m images. SPIRE band images are not included in the creation of the TIR map because it would drastically degrade the resolution of the resulting map. Different UV-to-optical/NIR colours provide good estimators of the sSFR on global or resolved scales. When available, we prefer the ${\text{FUV}}-r$ colour, as the $r$ band image usually combines a high resolution with a high signal-to-noise. Based on this colour map, we create a sSFR map (which is smoothed by convolving it with a kernel with twice the FUV FWHM to eliminate noise), and this map is used in combination with the TIR and FUV maps to determine the FUV attenuation and intrinsic (unattenuated) FUV emission maps.

Subsequently we correct this intrinsic FUV map for the contribution from old stars, by subtracting a scaled version of the 3.6~$\mu$m image from this map. We determine the best scaling factor by the assumption that the contribution from star formation in the bulge-dominated region is minimal. We hence define an elliptical region that corresponds to the bulge-dominated region of the galaxy, and determine, for each scaling factor, the relative number of pixels that have a negative value in that region of the subtracted map. We take 10\% as the upper limit of negative pixels that the corrected map can have so that the subtraction still makes physically sense (assuming 10\% negative values are largely due to noise). After the subtraction, a smoothing is applied to interpolate over the negative values. We refer, for instance, to \citet{Leroy2008, Rahman2011, Cortese2012, Ford2013, Casasola2017} for a similar analysis. 

This intrinsic young non-ionising stellar population FUV map is de-projected to a 3D disc model using the {\texttt{ReadFitsGeometry}} routines in SKIRT, in the same way as discussed for the old stellar population map. Following \citet{DL14} we use the same scale height for this population as for the dust, i.e.\ half of the scale height of the old stellar population (see Sect.~{\ref{Dustdisc.sec}}).

Finally, we assign an intrinsic spectral energy distribution and a normalisation to this component. We again use a \citet{BruzualCharlot2003} SSP model, with the same metallicity as for the old stars, but now with an age of 100~Myr. The normalisation, i.e., the total luminosity, is left as a free parameter in the model.

The resulting young non-ionising stellar disc of M~81 is shown in Fig.~\ref{fig:maps}.

\subsubsection{The young ionising stellar disc}
\label{sssec:ionising}

The presence of very young, partially obscured stars can be traced by H$\alpha$ emission, which, unfortunately, is again affected by attenuation. Using a sample of 33 nearby galaxies, \citet{Calzetti2007} present a prescription to create an extinction-corrected H$\alpha$ map, based on the observed H$\alpha$ map and the MIPS 24 $\mu$m map, 
\begin{equation} \label{eq:yi}
S_{\text{H$\alpha$,\,corr}} = S_{\text{H$\alpha$}} + 0.031\,S_{24} 
\end{equation}
We first correct the observed 24 $\mu$m image for a contribution from the old stellar population that may also emit at these wavelengths, using a similar approach as discussed for the contribution of old stars to the FUV emission. We use this corrected 24 $\mu$m image and the H$\alpha$ image, convolved and re-binned to the same resolution, and use the equation above to obtain the map of the young ionising stellar component, shown in Fig.~\ref{fig:maps}. This map is again de-projected in the same way, with a scale height that is half the value of the scale height of the young non-ionising stellar population, i.e., a quarter that of the old stellar disc \citep{V17}.

For the young ionising stellar population, we adopt the SED templates for obscured star formation from \citet{Groves2008}, based on 1D simulations of Starburst99 SSP models \citep{Leitherer1999} and the photoionisation modelling code MAPPINGS III \citep{Groves2004}. The result of their modelling is a suite of templates defined on a five-dimensional grid of the metallicity of the gas ($Z$), the mean cluster mass ($M_{\text{cl}}$), the compactness of the clusters ($C$), the pressure of the surrounding ISM ($P_0$), and a fraction controlling the optical depth of the gas and dust for the UV photons ($f_{\text{PDR}}$). The normalisation of the templates scales with the star formation rate (SFR). We used the following parameters as our default values: $Z=0.03$, $M_{\text{cl}} = 10^5~M_{\odot}$, $\log{C} = 6$, $P_0/k = 10^6~{\text{cm}}^3~{\text{K}}$, and $f_{\text{PDR}} = 0.2$. Similar values have been used by \citet{DL14} and \citet{V17} in their radiative transfer modelling of M~51 and M~31, and in other studies that used these MAPPINGS III SEDs as templates for star forming regions in galaxy-wide simulations \citep[e.g.,][]{Camps2016, Trayford2017, Camps2018}. All template SEDs are implemented in SKIRT for immediate use in combination with any kind of geometry. 

Finally, the normalisation of the young ionising stellar disc component, i.e. the total luminosity, is a free parameter in the modelling.

\subsubsection{The diffuse dust disc}
\label{Dustdisc.sec}

The final component to be considered is the dust disc. A map of the dust surface mass density can be obtained directly from the FUV attenuation \citep{DL14}. For M~81, the dust map, shown in Fig.~\ref{fig:maps}, shows some interesting features. Two clear spiral arms are visible, coinciding with those found for the young stellar populations, yet the center of the galaxy is mostly devoid of dust. Such a depression of the dust surface density is found in many DustPedia galaxies \citep{2019A&A...622A.132M}. A secondary spiral pattern is visible in the inner region, as already noted by \citet{Connolly1972} and \citet{Devereux1997}.

The dust map is de-projected by SKIRT to a 3D dust distribution, assuming a dust scale height that is half that for the old stellar population. This value is driven by the results of detailed radiative transfer modelling of edge-on spiral galaxies \citep{Xilouris1999, Bianchi2007, DeGeyter2014, Mosenkov2018}.

For the dust composition, we use the THEMIS\footnote{\url{https://www.ias.u-psud.fr/themis/}} dust mix appropriate for the diffuse ISM \citep{Jones2017}. The model is calibrated to reproduce the dust emission and extinction with a mixture of amorphous carbonaceous grains and silicate grains. The total dust mass, $M_{\text{dust}}$, which acts as the normalisation constant for the dust distribution, is a free parameter of the model. 

Table~\ref{tab:maps} gives an overview of the different components in the RT model that are constructed based on a 2D map and de-projected. Their intrinsic resolutions, depending on the used observational images, are also listed in this table.

\begin{table*}[]
\centering
\caption{Summary of the maps used for the radiative transfer model and their properties, as well as the observed images that have used to produce each map. The image with the lowest resolution, thus determining the resolution of the map, is indicated in bold.}
\label{tab:maps}
\begin{tabular}{llcc}
\hline
component        & images used  & pixel scale & FWHM \\
 & & [arcsec] & [arcsec] \\ \hline
old stellar disc & \textbf{IRAC~3.6~$\mu$m} & 0.75 & 1.9  \\
young non-ionising disc      & {\textit{GALEX}}~FUV, IRAC~3.6~$\mu$m, SDSS~$r$, MIPS~24~$\mu$m, PACS~70~$\mu$m, \textbf{PACS~160~$\mu$m} & 4.0                 & 11.2    \\
young ionising disc   & \textbf{MIPS~24~$\mu$m}, IRAC~3.6~$\mu$m, H$\alpha$ & 2.03 & 6.43 \\
dust             & SDDS~{\textit{r}}, {\textit{GALEX}}~FUV, MIPS~24~$\mu$m, PACS~70~$\mu$m, \textbf{PACS~160~$\mu$m}  & 4.0 & 11.2 \\       
\hline
\end{tabular}
\end{table*}

\subsection{Set up of the SKIRT simulations}
\label{ssec:skirt}

Based on the input maps and a minimum set of additional input parameters, our PTS pipeline creates the maps and 3D components necessary for the SKIRT radiative transfer modelling, and it also automatically generates the SKIRT parameter file \citep[for details on and an example of a SKIRT parameter file, see][]{CampsBaes2015}. Before a simulation can be run, a number of additional settings need to be chosen.

The spatial resolution of a SKIRT radiative transfer model is defined by the discretisation of the dust component in the so-called dust grid. The density and other dust properties are therefore constant within each unit of the grid (dust cell), and during the SKIRT simulation, all photon packages are propagated across this grid. The stellar emission is sampled on the input resolution of the stellar component maps. To accommodate the complexity of the structure in the dust distribution and limit the required computing resources as much as possible, we use the built-in binary tree grid structure in SKIRT \citep{Saftly2014}, which constructs the grid in a hierarchical way by splitting cells alternately along the three axis planes. The minimum and maximum subdivision level are chosen to be 6 and 36, respectively, and we set the stopping criterion for individual cells to the point where they contain less than $5 \times 10^{-7}$ of the total dust mass \citep{Saftly2013}. This threshold is similar to values used in the previous SKIRT simulations of spiral galaxies \citep{DL14, Saftly2015, Camps2016, Camps2018, V17, Trayford2017}.

For the specific case of M~81, the total extent of the dust grid is 26~kpc in the $x$ and $y$ direction, and 3.32~kpc in the vertical direction. The maximum level of 36 subdivisions corresponds to a maximum spatial resolution of $26~{\text{kpc}} / 2^{12} = 6.36~{\text{pc}}$, or 0.36~arcsec at the distance of M~81. As a result of the mass fraction criterion, the subdivision was truncated after level 27, such that the effective resolution was 50.9~pc, or equivalently 2.84~arcsec. In total, SKIRT generated a dust grid with 2.9 million cells; the leaves of a hierarchical tree consisting of 5.8 million individual nodes.

Another important aspect of any SKIRT simulation is the choice of the wavelength grid. SKIRT uses a fixed wavelength grid, in the sense that only photon packages with specific wavelengths are used in the simulation. For the choice of the specific wavelength grid in our modelling, several criteria are taken into account. On the one hand, the number of wavelengths needs to be restricted, as the run time of a single simulation scales roughly linearly with the number of wavelengths. On the other hand, the different input stellar SEDs and the expected dust emission SEDs need to be properly covered, sharp emission and absorption features need to be resolved, and the different broadband filters need to be properly sampled to allow for a proper convolution with the filter curves. A more detailed description of the wavelength grid construction is given in Appendix~\ref{sec:appendix_wavelengthgrid}. Balancing accuracy and computational feasibility, we settle on 250 wavelength points distributed in a non-uniform way over the entire UV-mm wavelength range. 

Finally, we simulate data cubes of the observed radiation. For the M~81 model, we choose the characteristics of the data cubes with a pixel scale of 4~arcsec or 71.6~pc, and a field of view of $17.9 \times 31~\text{kpc}$. This yields data cubes of $250 \times 433 \times 250 = 27.1$~million individual pixels. The instrument is placed at the galaxy distance of 3.69~Mpc, with the inclination angle of 57.8\textdegree\ and position angle of 66.3\textdegree\ also assumed for the de-projection.

\subsection{$\chi^2$ optimisation}
\label{ssec:optimisation}

The best-fitting values for the remaining free parameters in the RT model are determined through a $\chi^2$ optimisation procedure. In the setup described above, we have three free parameters: the dust mass $M_{\text{dust}}$, the FUV luminosity of the young non-ionising stellar disc $L_{{\text{yni}},\,{\text{FUV}}}$, and the FUV luminosity of the young ionising stellar disc $L_{{\text{yi}},\,{\text{FUV}}}$. Note that the luminosity of the bulge and the old stellar disc are not free parameters in the model, because their normalisation is based on the 3.6~$\mu$m emission which is relatively free of dust attenuation or contamination by young stellar populations. It is obviously possible to adapt this scheme and consider these luminosities as additional free parameters, or to include other free parameters such as the galaxy inclination or the dust optical properties if that would be required or interesting for the specific galaxy or the scientific questions to be addressed.

Initial guesses for the free parameters are determined through SED fitting results from \citep{2019A&A...624A..80N}, performed with the CIGALE \citep{2019A&A...622A.103B}. For most of the nearby galaxies within the DustPedia sample, CIGALE SED fitting has been performed, and the results are available on the DustPedia archive. For this purpose, the THEMIS dust model was introduced into CIGALE, a delayed and truncated SFH has been assumed \citep{Ciesla2016}, and for the SSPs the \citet{BruzualCharlot2003} spectra are used. Various parameters are derived from the best fitting stellar and dust population mix, including the SFR, the stellar and dust mass, the FUV attenuation and the temperature of the dust. We convert this SFR into a spectral luminosity for the young ionising stellar component by generating the \citet{Groves2008} spectrum with the appropriate model parameters and scale it to the obtained star formation rate. Subsequently, we take the observed FUV luminosity, correct it with the FUV attenuation value found by CIGALE, and subtract the found FUV luminosity of the young ionising population to get the intrinsic luminosity of the young non-ionising stellar component. Finally, the dust mass from CIGALE is directly used as our initial guess for the dust mass in the SKIRT modelling. 

For the case of M~81, the total dust mass according to this CIGALE fit, and hence the initial guess for our SKIRT modelling, is

\begin{equation}
M_{\text{dust}}^{\text{init}} = 9.61 \times 10^6~M_\odot
\end{equation}

The SFR obtained for M~81 is 0.351~$M_\odot$~yr$^{-1}$, and applying the recipes as described above yield the following initial guesses for the FUV luminosities of the young ionising and non-ionising stellar discs,

\begin{gather}
L_{{\text{yi}},\,{\text{FUV}}}^{\text{init}} = 1.59 \times 10^{36}~{\text{W}}~\mu{\text{m}}^{-1} \\
L_{{\text{yni}},\,{\text{FUV}}}^{\text{init}} = 3.18 \times 10^{36}~{\text{W}}~\mu{\text{m}}^{-1}
\end{gather}

We now need to optimise our model parameters going from the initial guess to a set that best represents the observations. As these fully-3D panchromatic Monte Carlo radiative transfer simulations are computationally demanding, it is important to limit the total number of simulations. Because we restrict the number of free parameters in our optimisation to only three, the problem at hand is still conveniently solvable with a brute-force parameter grid evaluation method. We follow a two-step approach where we first test the parameter space around our initial guess with faster simulations, and then refine the parameters with detailed simulations.

We determine the best fitting radiative transfer model parameters evaluating the $\chi^2$ metric. For the first, exploratory grid, we compare the global fluxes of the model to the observed integrated SED of M~81:
\begin{equation}
\chi^2 = \sum_X \mathpzc{w}_X 
\left( \frac{F_X^\text{sim} - F_X^\text{obs}}{\sigma_X^\text{obs}} \right)^2
\label{eq:chi2global}
\end{equation}
where $F_X^\text{sim}$, $F_X^\text{obs}$ and $\sigma_X^\text{obs}$ are the global mock flux density, observed flux density and observed error corresponding to band $X$, respectively. The differential weight factors $\mathpzc{w}_X$ are chosen such that each wavelength regime has an equal importance \citep[see also][]{V17}. We define six different regimes (ionising, young, old, mix, aromatic, and thermal dust). Each of these regimes has been chosen to reflect the presumed dominating contribution of a particular model component, and thus the regime where this component is most likely to be best constrained. We assign an equal weight to each of the regimes since we want all model components to be equally well constrained, and determine the weight factor for the filters in a regime based on their count\footnote{We select the filters from the {\textit{GALEX}}, SDSS, {\textit{Spitzer}}, {\textit{WISE}} and {\textit{Herschel}} telescopes, but also we avoid filters that are too close in effective wavelength. In particular, we generally choose IRAC~3.6~$\mu$m over {\textit{WISE}}~3.4~$\mu$m, IRAC~4.5~$\mu$m over {\textit{WISE}}~4.6~$\mu$m, and MIPS~24~$\mu$m over {\textit{WISE}}~22~$\mu$m. This yields a set of 18 filters to be used for the fitting.}.

For the second, detailed grid, we compute the $\chi^2$ metric based on the pixel-by-pixel difference of observed versus mock images in each band:
\begin{equation}
\chi^2 = \sum_X \sum_{p} 
\mathpzc{w}_X 
\left[ \frac{\mu_{X}^\text{sim}(p) - \mu_{X}^\text{obs}(p)}{\sigma_{X}^\text{obs}(p)} \right]^2 
\label{eq:chi2local}
\end{equation}
The first summation covers all filters $X$ for which data are available, the second sum loops over all pixels $p$ in the image corresponding to band $X$. The quantities $\mu_{X}^\text{sim}(p)$, $\mu_{X}^\text{obs}(p)$ and $\sigma_{X}^\text{obs}(p)$ are the mock surface density, observed surface density and observed error in pixel $p$ of band $X$ respectively. Finally, $\mathpzc{w}_X$ is again a differential weight factor given to the band $X$, as in Eq.~\ref{eq:chi2global}. The choice for a local $\chi^2$ can be computationally demanding. Generating mock images with sufficiently high signal-to-noise in all pixels and in all bands requires for each model takes considerable computation time. The average CPU time in this phase of the M~81 modelling was 311h for each simulation.\footnote{All simulations together consumed around 107,000 CPU hours.} Depending on the level of detail and available resources, one could also fall back to the global $\chi^2$ in this step. This was done for M~31, for example, which is more than seven times larger than M~81 \citep[see][]{V17}. We note that this will become less of a problem as soon as the recently released new version of SKIRT (SKIRT 9: Camps \& Baes, submitted) has been fully validated. In this new version of the radiative transfer code, mock broadband images are directly generated as part of the simulation. Instead of producing large data cubes with many of small wavelength bins, the relevant photon packages of a particular band are captured directly and stored in a much smaller data cube of broadband images. This is a more time and memory efficient way of making mock images, which can subsequently be used to minimise the $\chi^2$ per pixel. The reduced $\chi^2$ values are converted into probabilities as $\exp(-\tfrac12\,\chi^2)$, and the probabilities of all models with a certain parameter value are summed to yield the probability of that particular parameter value.

The exploratory grid as a test of our initial guess is set as a coarse but broad grid in the $(M_{\text{dust}}, L_{{\text{yni}},\,{\text{FUV}}}, L_{{\text{yi}},\,{\text{FUV}}})$ parameter space. We consider five grid points for the dust mass and for the FUV luminosity of the young non-ionising disc, both on a logarithmic grid. Dust masses are generally reasonably well constrained by SED fitting, so we choose a range for the dust mass parameter of just one order of  magnitude around the initial value. The young non-ionising stellar disc FUV luminosity is harder to constrain, so we choose two orders of magnitude for this parameter. Based on the analysis for M~31 \citep{V17}, we assume that the normalisation of the young ionising stellar disc is generally even harder to constrain, so we conservatively use a broader grid of seven points for this parameter, spanning three orders of magnitude around the initial guess. Still, the ranges are chosen with the idea that one can always expand them if borders are hit. In total, the first batch contains 175 radiative transfer simulations.

As indicated above, we speed up the evaluation in this step by reducing the output detail (the simulation's internal 3D resolution remains the same). For the simulations in this exploratory grid, we generate a lower-resolution wavelength grid of just 115 wavelengths, abandoning the requirement of spectral convolution and thus only including the effective wavelength of each filter for a direct comparison to observation. We also do not create the full data cube for each simulation but let SKIRT create a global simulated SED. In this step, we use $10^6$ photon packages per wavelength, which is sufficient for generating SEDs. These adaptations provide almost an order of magnitude in speedup, but still 34 hours of CPU time were required per simulation.

Based on the results of this initial grid of models, we create a finer grid centred around the expected values of the free parameters. This time we use 7 grid points in each of the dimensions of the $(M_{\text{dust}}, L_{{\text{yni}},\,{\text{FUV}}}, L_{{\text{yi}},\,{\text{FUV}}})$ parameter space, resulting in 343 additional simulations. In these simulations, we use five times as many photon packages and employ the full high-resolution wavelength grid. We create a data cube for each simulation and spectrally convolve this cube to create a mock image for each observed filter. We subsequently re-bin these images to the coordinate system of the corresponding observed image and clip them to the same field-of-view. We can thus evaluate the $\chi^2$ measure on a per-pixel basis (Eq.~\ref{eq:chi2local}). This large computational effort is preferred over the minimisation of a global $\chi^2$ (Eq.~\ref{eq:chi2global}) because it also captures local variations between model and observation. The impact of using a global $\chi^2$ is discussed in Appendix~\ref{app:localglobal}.

\subsection{Summary of the modelling steps}

\begin{figure*}[t!]
\centering
\includegraphics[width=0.99\textwidth]{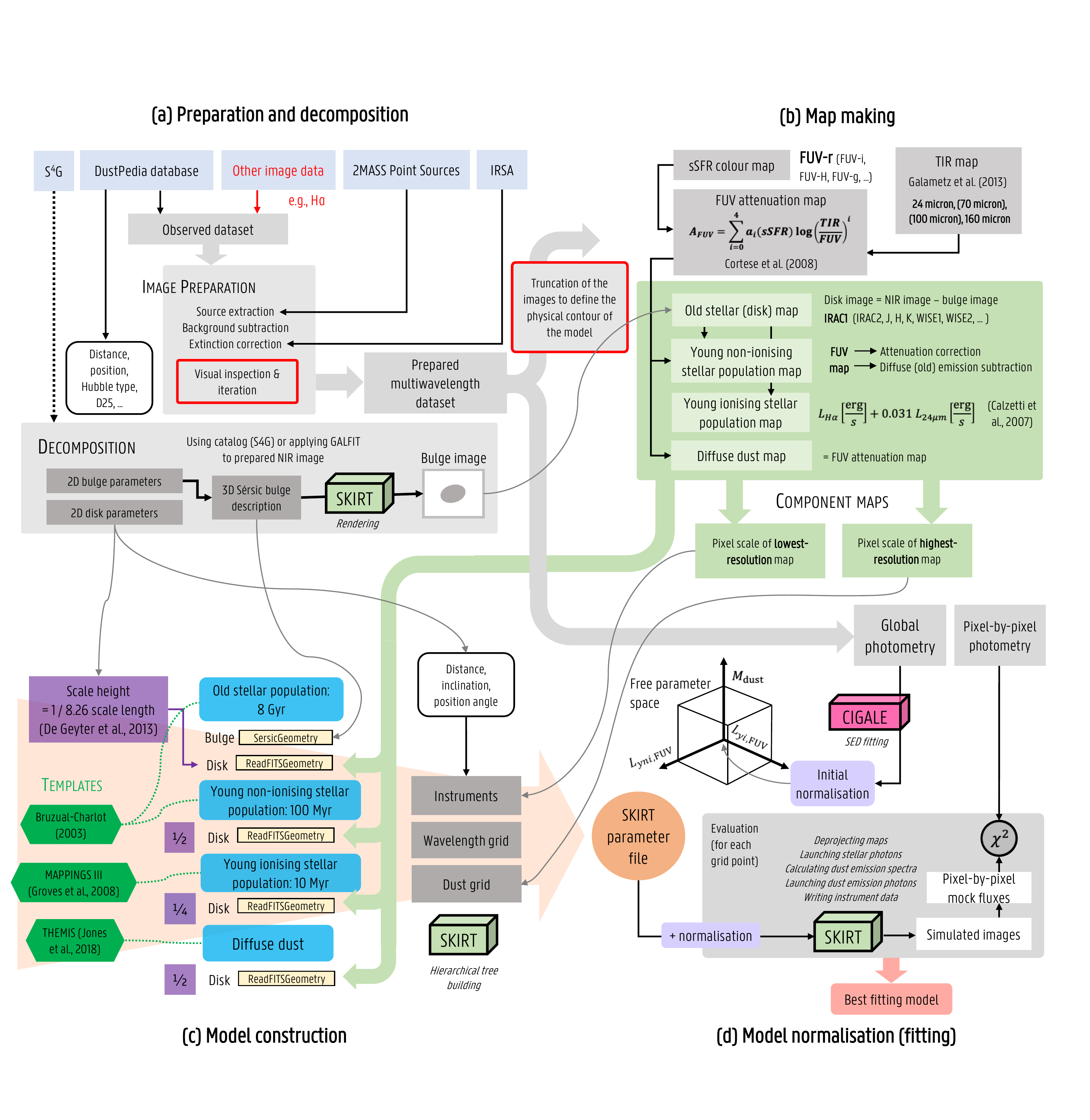}
\caption{The semi-automatic modelling approach, as implemented in the Python code PTS. Indicated in red are steps that require manual intervention. At the top of the diagram, the different input sources are displayed. These include the DustPedia archive, but also other image data can be included in the modelling environment. The images are prepared uniformly by the pipeline, using the 2MASS point sources catalog and IRSA service for dust extinction. If {\textit{Spitzer}}~3.6~$\mu$m or 4.5~$\mu$m data is available a 2D decomposition is retrieved from the S$^4$G catalog, else custom decomposition parameters can be specified. The prepared images are used to obtain a map of the different stellar and disc components based on well-established prescriptions. Crucial in the map making is the FUV attenuation map, which in turn is based on a sSFR colour map (the preferential colour is indicated in bold, though other colours can be used depending on the availability and quality of the data) and a TIR map (created by combining multiple MIR/FIR bands with a consideration of the resolution). The maps and 3D bulge description define the geometry of the RT model, together with appropriate scale heights for each of the disc components. The RT model is further defined by a wavelength grid, an instrument setup, and a dust grid. Given this basic model, SKIRT is used to determine the best normalisations of stellar and dust components. For more detail, we refer the reader to the main text in Section \ref{sec:approach}.}
\label{fig:framework}
\end{figure*}

Fig.~\ref{fig:framework} summarises outlines the general 3D radiative transfer modelling approach we have adopted and will also apply for future work. It shows the key steps that are involved in the modelling, such as the image preparation, the map making, the decomposition, and the fitting steps. 

\section{Results}
\label{sec:results}

\subsection{Model selection and properties}

\begin{figure*}[t!]
\centering
\includegraphics[width=0.99\textwidth]{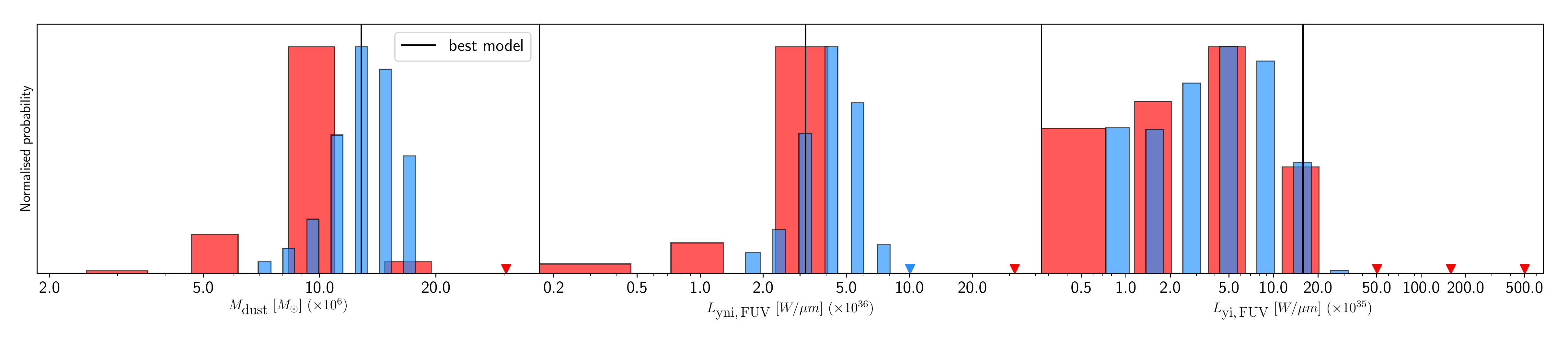}
\caption{Probability distributions of the RT models for the first batch (red) and the second batch of simulations (blue). From left to right: the diffuse dust mass, the FUV luminosity of the young non-ionising disc, and the FUV luminosity of the young ionising stellar disc. For the second run, the values of the overall best-fitting simulation are indicated by the vertical lines.}
\label{fig:pdfs}
\end{figure*}

The probability distributions from the 175 simulations of the first set of models for M~81 are shown in red in Fig.~{\ref{fig:pdfs}}. For the dust mass and the FUV luminosity of the young non-ionising stellar disc, the expected value is the initial guess value. For the FUV luminosity of the young ionising stellar disc, the grid point with the highest probability is $5.02 \times 10^{35} \, \text{W}/\mu\text{m}$, one grid point below the initial guess value. This shows that models deviating significantly from the initial guess values are punished by a higher global $\chi^2$. After this test grid we can be confident of our setup and thus run the more detailed simulations of the more narrow, but better sampled parameter grid. The resulting probability distributions are over-plotted in blue in Fig.~{\ref{fig:pdfs}}. Also indicated are the parameter values of the best fitting simulation (lowest local $\chi^2$). The most probable values (peak of the blue distributions) of the  parameters are

\begin{gather}
M_{\text{dust}} = \left( 1.28 \pm 0.27 \right) \times 10^7~M_\odot \\
L_{\text{yni,\,FUV}} = \left( 4.24 \pm 1.43 \right) \times 10^{36}~{\text{W}}~\mu{\text{m}}^{-1} \\
L_{\text{yi,\,FUV}} = \left( 5.02 \pm 4.78 \right) \times 10^{35}~{\text{W}}~\mu{\text{m}}^{-1}
\end{gather}
The most likely radiative transfer model for M~81, i.e.\ the model with the lowest local $\chi^2$, has parameters
\begin{gather}
M_{\text{dust}} = 1.28 \times 10^7~M_\odot \\
L_{\text{yni,\,FUV}} = 3.18 \times 10^{36}~{\text{W}}~\mu{\text{m}}^{-1} \\
L_{\text{yi,\,FUV}} = 1.59 \times 10^{36}~{\text{W}}~\mu{\text{m}}^{-1}
\end{gather}
The model with these parameter values is the one we use in the rest of the analysis.

\subsection{Spectral energy distribution}
\label{SEDComparison.sec}

\begin{figure*}[h]
\centering
\includegraphics[width=0.99\textwidth]{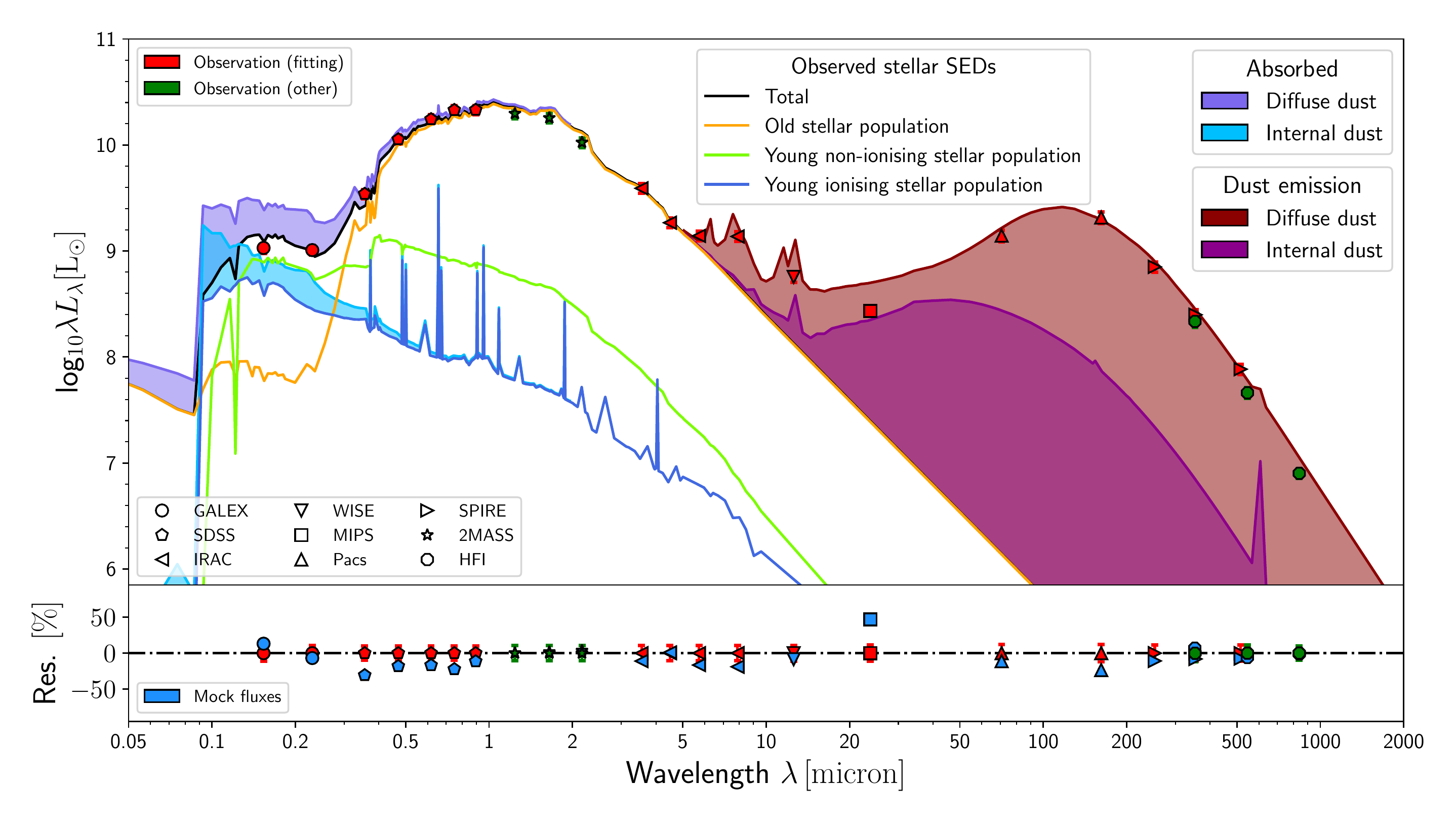}
\caption{Top panel: the panchromatic SED corresponding to the SKIRT radiative transfer model for M~81. The total observed stellar energy output is indicated with a black line and the total dust emission, when added to the stellar spectrum, follows the red line. The area shaded in purple above the observed stellar spectrum represents the radiation that is absorbed by the diffuse dust, thus the purple line represents the intrinsic stellar spectrum. Yellow, green, and dark blue lines represent the contribution of the old, the young non-ionising, and the young ionising stellar populations to the observed stellar spectrum, respectively. The area shaded in light blue above the young ionising population curve indicates the absorbed energy by internal (subgrid) dust. The area shaded in magenta between 20 and 500 $\mu$m indicates the energy distribution emitted by this internal dust. Bottom panel: A comparison between the observed fluxes and the simulated SED, as relative residuals with the observations as reference. Observed fluxes are shown as red and green points, depending whether or not they were used in the fitting procedure. The blue points show the mock fluxes derived from the simulated SED.}
\label{fig:seds}
\end{figure*}

The simulated SED of the best fitting model is shown in Fig.~{\ref{fig:seds}}, along with the observed photometry derived from the prepared images. The total model SED is represented by the black line starting from the left side of the spectrum, taken over by the red line where the dust emission adds to the stellar part of the model SED on the longer wavelength side of the panel. Overall, the simulation reproduces the observations notably well. A direct comparison on a global scale is shown in the bottom panel of Fig.~{\ref{fig:seds}}, where the reference is the observation (red points for the bands included in the fitting, green points for other observed fluxes). The mock fluxes (blue points) are derived from the simulated luminosity's by spectrally convolving them with the appropriate filter transmission curves and applying the same aperture photometry as we performed on the observations. Residual percentages are defined as (model-observation)/observation.

The most prominent deviation of the model from the observation is seen for the MIPS $24 \, \mu$m band($50\%$). This band lies close to the peak of the hot dust component related to the ionising stellar populations. It seems that, on a global scale, our model overestimates this contribution. It is important to note that models with lower $L_{\text{yi,\,FUV}}$ did not provide a better fit to the data. While reducing the MIPS $24 \, \mu$m emission, they lower the MIR continuum entirely and thus producing a higher $\chi2$. In comparison, the WISE $22 \, \mu$m flux (not shown, but see Table~\ref{tab:fluxes}) is $10\%$ higher than the MIPS $24 \, \mu$m emission. Part of this discrepancy may also be related to instrumental uncertainty in the MIR. The residuals for the other bands are at maximum $25\%$ and usually lower. The 2MASS and \textit{Planck} fluxes were not included for the fitting procedure because of significant large-scale sky noise and poor spatial resolution, respectively. Nevertheless these also agree well with the model (5\%).

\subsection{Image comparison}
\label{ImageComparison.sec}

\begin{figure*}[t!]
\centering
\includegraphics[width=\textwidth]{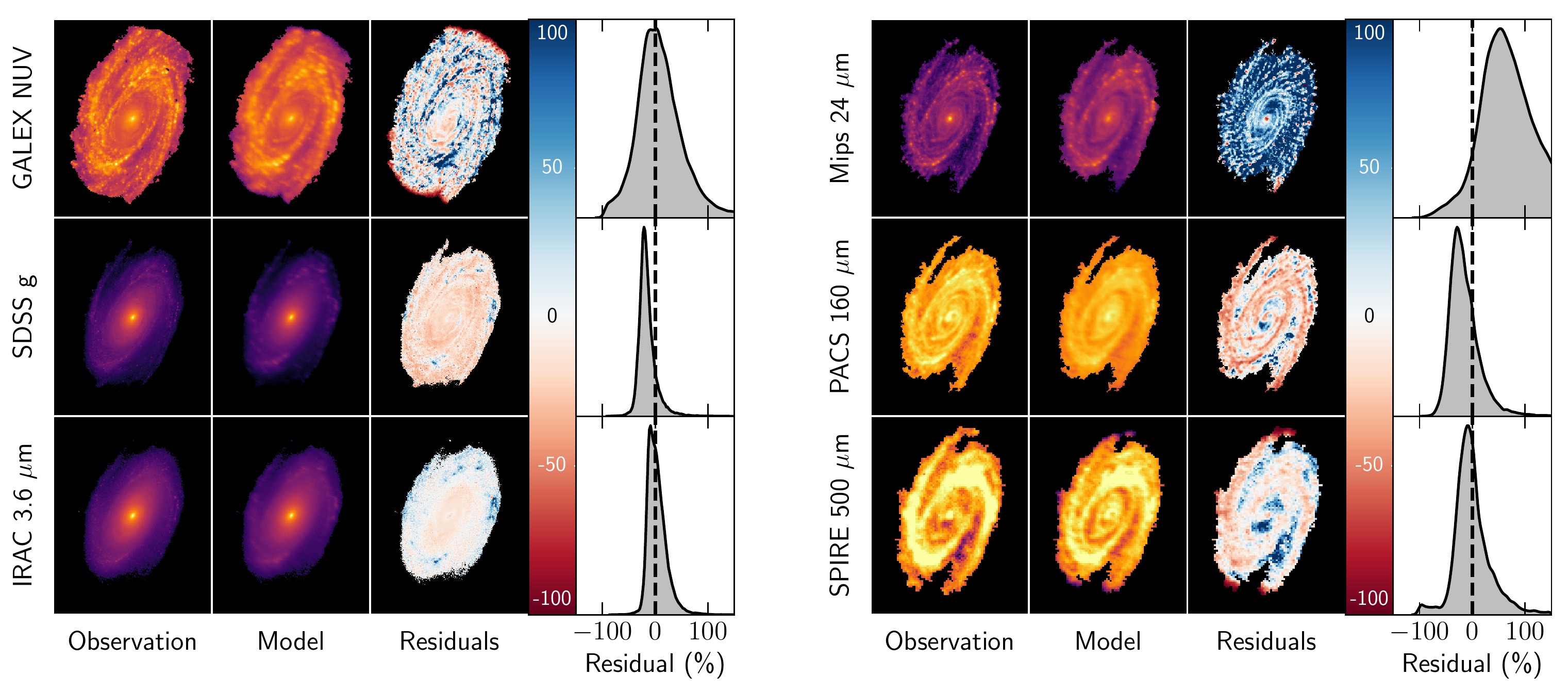}
\caption{Comparison between observed and simulated images in 6 different wavebands. First column: observed images. Second column: mock images, created by spectrally convolving the simulated data cube. Third column: maps of the relative residuals between observed and mock images. Last column: distributions of the residual pixel values. The mock images are clipped to the same pixel mask as the observed images. The probability density functions of the residuals are calculated with a kernel density estimation (KDE).}
\label{fig:residuals}
\end{figure*}

\begin{figure}[t!]
\centering
\includegraphics[width=\columnwidth]{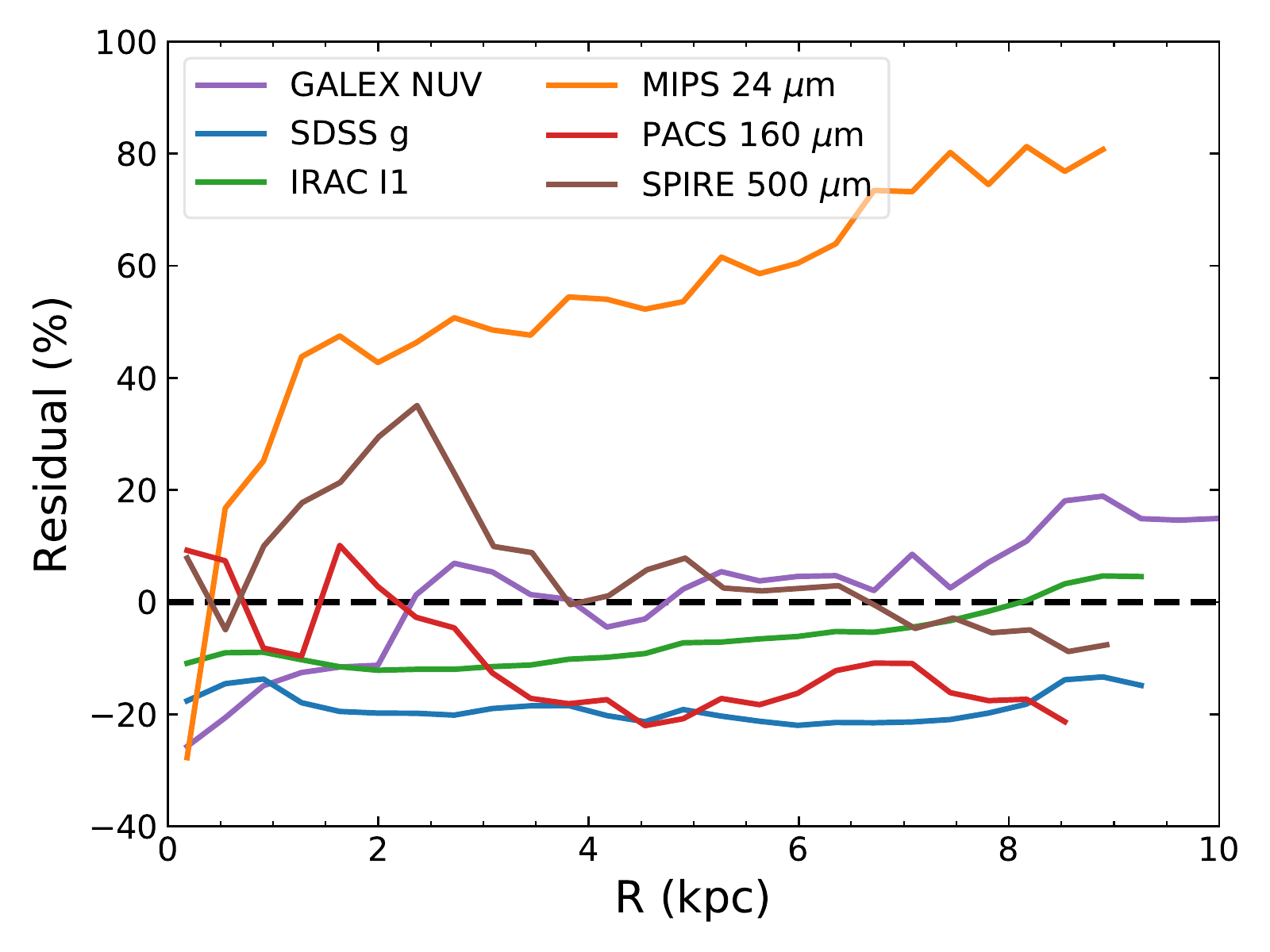}
\caption{Median residual versus galactocentric radius derived from the residual maps in Fig.~\ref{fig:residuals}. Each waveband is assigned with a different colour indicated in the upper left corner of the figure.}
\label{fig:residual_median_radial_profiles}
\end{figure}

Spectral convolution of the simulated data cubes for the 18 bands (used for fitting) gives rise to a set of mock images for every model of the second batch of the fitting procedure. A selection of such images, clipped to the observed high signal-to-noise mask, for the best fitting simulation is shown in Fig.~\ref{fig:residuals}. Here, the first column shows the observations and the second column the mock images. Because model and observation are generally difficult to discriminate, we highlight the small-scale relative differences through residual maps, shown in the third column. The last column shows the probability density distribution of the residual pixel values. In Fig.~\ref{fig:residual_median_radial_profiles}, we additionally show the corresponding radial profiles derived from the residual maps.

In the NUV band there is an overall good agreement with most deviant pixels well below $100\%$. The largest offsets occur in the inter-arm regions. Here the model over-estimates the flux as part of the applied deprojection of the 2D input maps. In the bulge the model slightly underestimates the UV flux. As such there's a rising trend of residual percentage with galactocentric radius. The pixel distribution is more narrow in the $g$ band, where model and observed image look very similar, giving rise to a flat radial profile. There is however a global offset of $\sim 25\%$ as already noted in the previous section. The main factors influencing the optical SED fit are the stellar populations and the dust attenuation. Since the offset in the $g$ band is uniform across the disc, the latter is unlikely responsible. Instead, this may be caused by insufficient resolution in the star formation history of our simulation (modeled by the three stellar population ages). A fourth component with an age between 100 Myr and 8 Gyr, or alternatively a stellar component with a continuous range of ages, could alleviate the offset in global flux while maintaining the smooth residual map.

In the NIR, there is a narrow residual distribution as well. The IRAC 3.6 $\mu$m pixels follow the observations closely. A weak radial trend is visible where low-luminosity disk pixels are are slightly overestimated and bulge pixels tend to be underestimated by the model. The good agreement in this band confirms our assumptions that the young stellar populations have no relevant contribution to the emission and that dust attenuation is insignificant here.

The situation is considerably worse in the MIPS 24 $\mu$m band, where the model typically generates pixel fluxes that are too high (see also  Sect.~\ref{SEDComparison.sec}). The corresponding radial profile in Fig.~\ref{fig:residual_median_radial_profiles} shows a steep rise from negative percentages in the centre to positive residuals beyond 1 kpc. The residuals keep increasing outwards, but at a slower rate. This band was used in combination with $H\alpha$ to construct the surface density map of the young ionising component (Eq.~\ref{eq:yi}). It is possible that this empirical prescription is introducing the radial trend in the residuals obvious in Fig.~\ref{fig:residual_median_radial_profiles}. In addition the emission template for the young ionising component has several tuneable parameters that can produce relatively higher 24 $\mu$m emission. We unfortunately do not have sufficient observational constraints to explore this vast parameter space. For consistency comparability, we adopted the same values as other radiative transfer model of this kind \citep{DL14, V17, 2019MNRAS.487.2753W}.

The only regions where the 24 $\mu$m flux is underestimated by the model are the very centre of M~81 and the clumpy star forming regions in the spiral arms. The latter is seen in other bands as well, yet mostly where star formation is prominent (for example in the UV). The origin of this blurring is twofold, partly showing the effects of de-projection and its assumptions, yet another important degradation in resolution comes from the fact that multiple bands are combined into the input maps of the model components. For the 24 micron band, for instance, the observed image has a resolution (FWHM) of 6.43 arcsec, but the simulated emission map is modulated by the dust map which has a resolution of 11.18 arcsec (as does the young non-ionising stellar map), even though this dust is heated by the young ionising population (FWHM = 6.43 arcsec) as well. As a result of our choice of retaining the maximal resolution for each component separately, the resolution of the resulting data cube is the combined effect of whichever input maps have the strongest imprint on any particular part of the wavelength spectrum.  

Fig.~\ref{fig:residuals}, furthermore shows the residuals of the PACS 160 micron and SPIRE 500 micron bands, for which the agreement is mostly within 50\% for individual pixels. At $160 \, \mu$m there is a small global offset, which disappears at longer wavelengths. The corresponding radial profiles correlate and are mostly flat. At around 2 kpc there is a small bump of positive residuals which is linked to an area of low signal-to-noise in the submm maps. The model again underestimates the flux in the spiral arms and overestimates the inter-arm regions. The effect is largest in these dust emission bands and is probably a combination of two effects. First, the projection effect was already mentioned. In this case both young stars and dust are smeared out slightly due to the deprojection, making the resulting dust emission less peaked in the arms. Secondly, it is possible that the FUV attenuation map, which sets the input dust surface density, yields a smoother dust distribution than in reality.  When this map is subsequently scaled (by our optimisation procedure) to match the observations, it can cause the observed large-scale asymmetry between arm and inter-arm regions. The $500 \, \mu$m band is relatively uncontaminated, tracing the bulk of the dust mass. Despite the discrepancies, the model still matches the observations within 50\%, which indicates that the FUV attenuation map is still a very good tracer of the dust column density. This importantly allows us to work at higher spatial resolutions compared to a dust mass map derived from \textit{Herschel} data alone.

We conclude from this comparison that the radiative transfer model for M~81 is represents the data well and is of sufficient quality to address the goal of this study: the relative contribution of young vs old stellar populations to the dust heating. At the same time we recognise that the best-fit model has some flaws. Most notably are the inevitable projection effect when going from 2D input maps to 3D density distributions. In addition the model produces too much $24 \, \mu$m emission to match the data. While there is some uncertainty in the flux calibration of the data itself (compared to e.g. WISE), the amount of hot dust is definitely overestimated by the model. We keep these caveats in mind when analysing the dust heating in Sect.~\ref{ssec:heating_fraction}.

\subsection{Dust absorption}

To investigate how different sources of radiation heat the dust in the galaxy, we have set up individual simulations that contain either just the old stellar population (bulge + disc), the young non-ionising stellar disc, or the young ionising stellar population. These models, all containing the original dust distribution, are run with the same high-resolution configuration as used for the second-batch simulations. As seen in Fig.~\ref{fig:seds}, the young ionising population (dark blue line) is clearly the least important, which does not exclude that they are significant or even dominating on local scales. The young non-ionising stellar population is responsible for the emission in the UV bands, while the old stars dominate in all of the optical and NIR wavebands. 

Globally, $1.73\times10^{36}$~W is absorbed and re-emitted from $1.53 \times 10^{37}$~W of intrinsic stellar radiation. This corresponds to a bolometric attenuation $f_{\text{abs}} = 11.3\%$. This number agrees very well with the value of 10.8\% found by \citet{2018A&A...620A.112B} for M~81, but it is significantly below the average value of 25\% for 40 DustPedia Sab galaxies of this type. There is, however, a large scatter in $f_{\text{abs}}$ for Sab spirals, with 1$\sigma$ dispersion of 19\%.

For only the diffuse dust, we find an absorption rate of $1.49\times10^{36}$~W (9.8\%). In the FUV band, 53\% of the stellar luminosity is absorbed by the total dust population. This is equivalent to $A_\text{FUV}=0.69$~mag. For the old stars, just 6.9\% of their bolometric stellar luminosity is absorbed and re-emitted by dust, 6\% of the bulge stellar energy and 7.8\% of the disc stellar energy. The young non-ionising stellar population has an absorption fraction of about 30\%.

The absorption fraction for the young ionising stellar component can be expressed in multiple ways. Firstly, We can consider just the internal dust, i.e. the dust clouds that are a subgrid component of which the emission is represented in the specific MAPPINGS template. This dust absorbs 33.8\% of the intrinsic stellar luminosity (29\% in the FUV band). Of the same intrinsic energy budget, 32.1\% is additionally absorbed by the diffuse dust component in the model (35.3\% in the FUV band). Taken together, dust reprocesses about 62\% of the young ionising stellar radiative energy, and about 64.4\% of the emitted FUV radiation. The young stellar populations together are subject to an absorption fraction of 41.5\% (diffuse dust) or 44\% (all dust).

\subsection{Dust heating fraction}
\label{ssec:heating_fraction}

We are interested in quantifying the fraction of heating by young stars, $f_\text{young}$, as it indicates which age components will be picked up by common SFR tracers. It is thus important to know how much of that light is used to heat dust. Knowing the absorbed luminosity in each dust cell, originating from the different stellar components in the model, we can define the fraction of dust heating that is attributed to young stars as follows:

\begin{equation}
\label{eq:fyoung}
f_\text{young} = \frac{L^\text{abs}_\text{young}}{L^\text{abs}_\text{total}},
\end{equation}

where, in this context, the young population combines the contributions from the non-ionising and ionising subpopulations. The luminosity absorbed from the young ionising stellar population must include the dust that is subgrid in the MAPPINGS templates, as opposed to the other stellar components in the model, which only heat the diffuse dust. We have calculated $f_{\text{young}}$ for each cell in the RT model by using the absorption information written out by the young non-ionising, ionising and total simulations. 

\begin{figure*}[t!]
\centering
\includegraphics[width=0.97\textwidth]{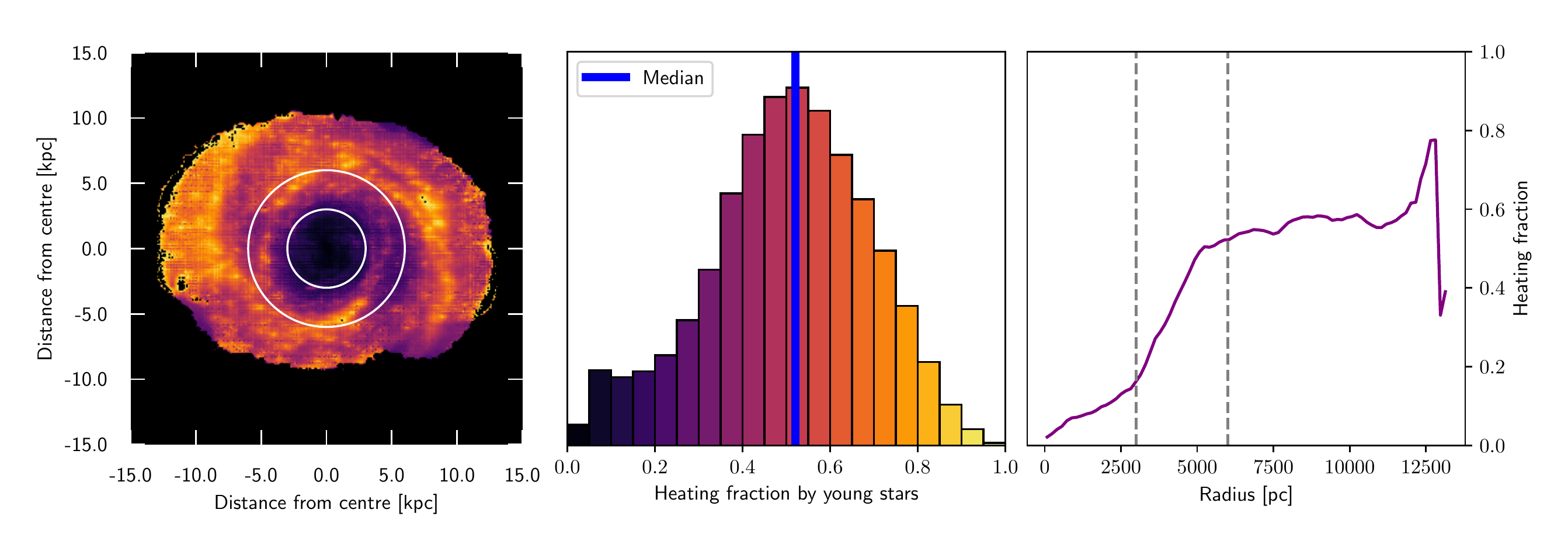}
\caption{Left panel:~Face-on map of $f_{\text{young}}$, the heating fraction by young stars, as obtained from the 3D dust cell data. White circles indicate the 3 kpc and 6 kpc radii. Central panel:~Distribution of the cell heating fractions, weighed by dust mass. Right panel:~The radial profile of $f_{\text{young}}$ as measured from the cell data.}
\label{fig:heating}
\end{figure*}

The left panel of Fig.~\ref{fig:heating} shows $f_{\text{young}}$ from a face-on view. The heating fraction by young stars goes from 0 to 1 with a colour scale that is depicted by the histogram in the central panel. This histogram shows the distribution of $f_{\text{young}}$ within dust cells, weighed by dust mass. We find that the bulk of dust mass in the model is heated by both young and old stars by about equal extent. However, an upturn in the lowest young heating fractions is noted, clearly corresponding to the center region of the galaxy as seen in the left plot. This region is dominantly heated by the old stellar bulge, and to lesser extent by the old stellar disc. Globally, 50.2\% of the heating (absorption) originates from the young stars, with lowest values of 0.5\% in the most central dust cells. Within the central region indicated by the inner white circle in Fig.~\ref{fig:heating}, with a radius of 3~kpc, the heating fraction by young stars is just 10.4\%.

We hereby confirm the findings of \citet{Bendo2012}, who conclude that the old stellar population is the dominant heating source within 3~kpc of the center of M~81, since star formation traced by H$\alpha$ is very low in this region but colour temperatures between 70~$\mu$m and 250~$\mu$m are elevated in this region. In the intermediate region between the two circles, $f_{\text{young}}$ is 42.4\%, and it is 56.2\% outside the outer circle with a radius of 6~kpc. We have also calculated $f_{\text{young}}$ in the mid-plane of the galaxy, and found there are only minimal differences with the face-on heating map. In fact, we found a monotonic decrease of $f_{\text{young}}$ with height at a rate of about 3.2\% per dust scale height. This indicates that $f_{\text{young}}$ varies only slightly within the vertical scale of the galaxy model, in contrast to the clear variation with galactocentric radius. We calculate the radial profile of $f_{\text{young}}$ by binning the dust cell data. In each radial bin, we calculate the average of $f_{\text{young}}$, again weighed by the dust cell masses. The resulting curve is shown in the right panel of Fig.~\ref{fig:heating}. Plotted as dashed grey lines are the inner and outer circles also plotted in the left panel of Fig.~\ref{fig:heating}.  

\begin{figure*}[t!]
\centering
\includegraphics[width=0.95\textwidth]{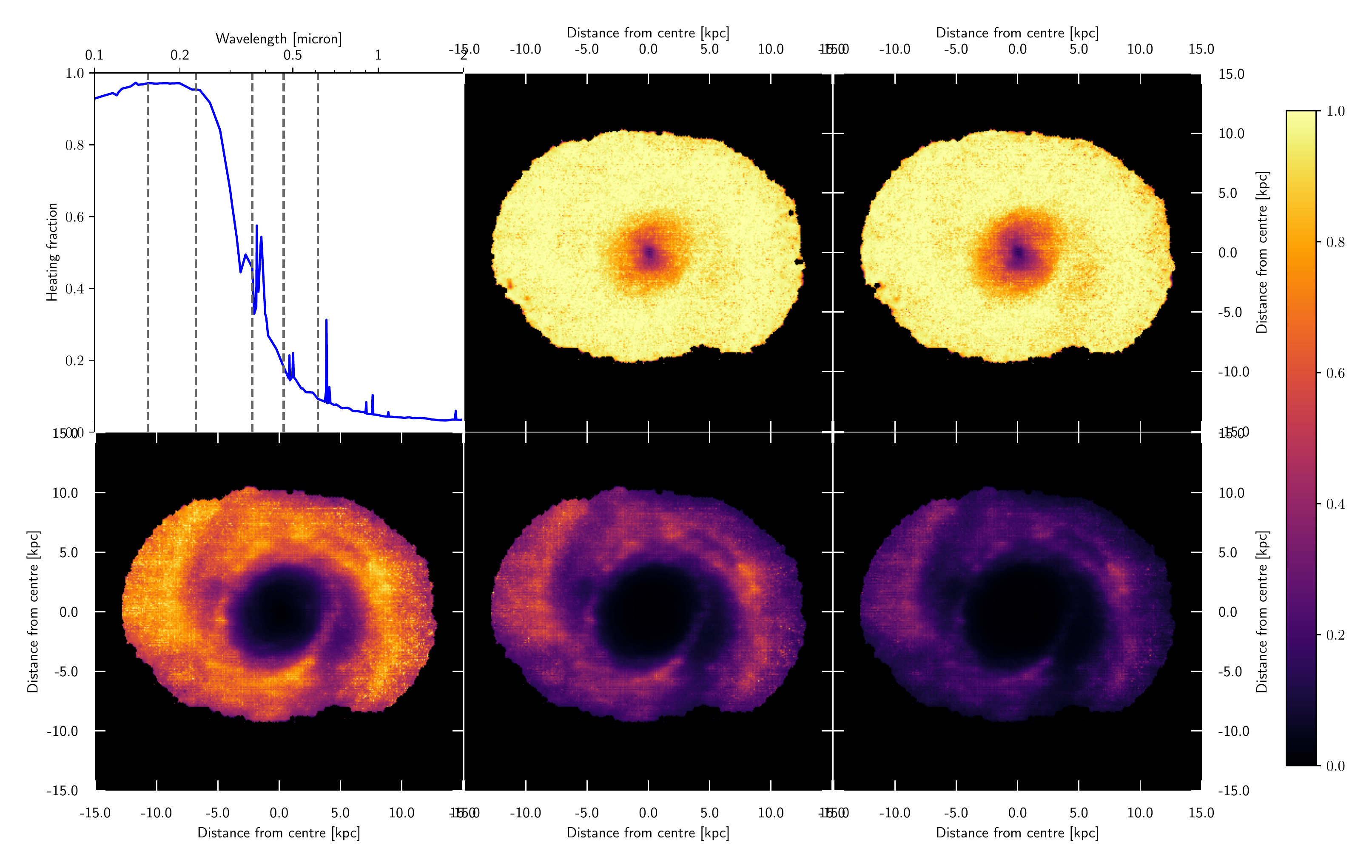}
\caption{The top left plot shows a curve of $f_{\text{young}}$ with respect to the wavelength at which the energy is absorbed. Five wavelengths have been indicated, corresponding to the {\textit{GALEX}}~FUV, {\textit{GALEX}}~NUV, SDSS~{\textit{u}}, SDSS~{\textit{g}}, and SDSS~{\textit{r}} effective wavelengths. For these filters, a projected map of $f_{\text{young}}$ in the dust cells is shown in the subsequent panels.}
\label{fig:spec_heating}
\end{figure*}

The absorption spectrum in each dust cell can also be used to express $f_{\text{young}}$ in each individual wavelength bin. The curve describing the wavelength-dependent heating fraction by young stars is plotted in the first panel of Fig.~\ref{fig:spec_heating}. Based on the 3D spectral heating fractions, we can make projected maps as in the previous section. Also in Fig.~\ref{fig:spec_heating}, we show maps of $f_{\text{young}}$ in the {\textit{GALEX}} FUV and NUV bands, and the SDSS~{\textit{u}}, {\textit{g}}, and {\textit{r}} bands. There is a dramatic shift from heating by young stars (in the FUV and NUV bands) to heating by old stars (from the SDSS~{\textit{g}} band onwards). In the SDSS~{\textit{u}} band, the fraction is around half on a global scale, but the disc is still predominantly heated by young stars. Notably, the most central region of the galaxy is always heated by old stars, over the entire wavelength range. 

\section{Discussion}
\label{sec:discussion}

\subsection{Correlation with sSFR}
\label{ssec:correlations}

The analysis above has shown that the main driver of dust heating greatly varies within the galaxy. On a spatially resolved scale, $f_{\text{young}}$ seems strongly linked to the presence of a dominant old stellar bulge in the center. While the star formation in the spiral arms dominates the heating locally, overall the old and young contributions in the disc balance out almost equally. Without a fully 3D radiative transfer model, obtaining these conclusions about the heating mechanisms would be hard, certainly considering the fact that in the global energy output of the galaxy model, the UV emission of the old stars is more than one magnitude lower than that of the young stars (see Fig.~\ref{fig:seds}). We have also shown that $f_{\text{young}}$ in the UV spectrum, and drastically decreases between 0.25 and 0.5~micron in our model. 

\begin{figure}[t]
\centering
\includegraphics[width=\columnwidth]{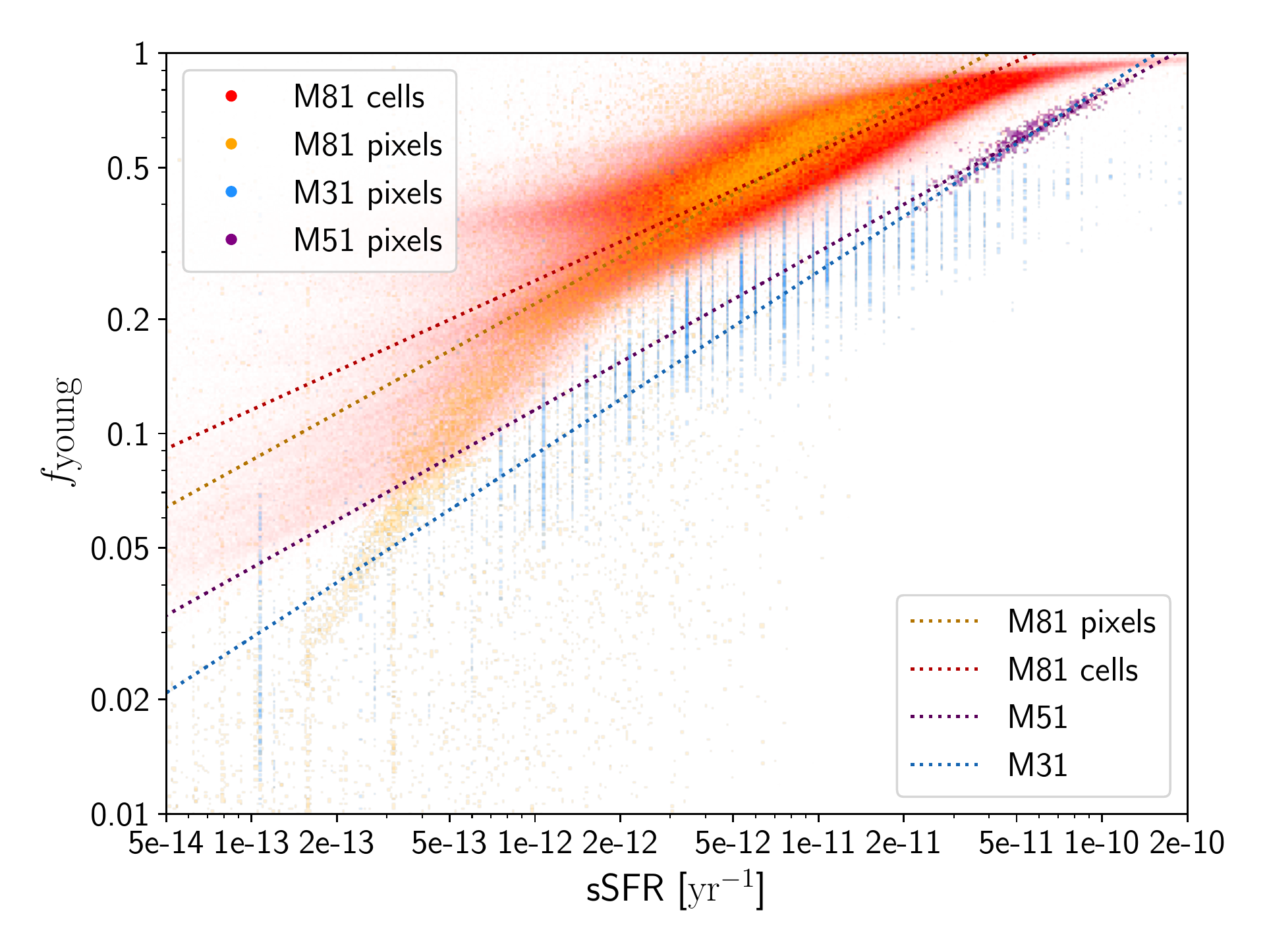}
\caption{Correlation between sSFR and $f_{\text{young}}$, shown as orange data points for the pixels and in red for the dust cells of the M~81 RT model. Blue and purple points are the values for the pixels of the M~31 \citep{V17} and M~51 model \citep{DL14}, respectively. Also shown are power-law fits to the data points in corresponding colours.}
\label{fig:fyoungssfr}
\end{figure}

\citet{DL14} and \citet{V17} have shown a link between $f_{\text{young}}$ and the specific star-formation rate (sSFR) in their radiative transfer models of M~51 and M~31 respectively. To compare with their results, we have calculated the star-formation rates in our model using a conversion from the intrinsic FUV luminosities of the young stars prescribed by \citet{Kennicutt2012}. To estimate the stellar masses, we have used the 3.6~$\mu$m luminosities, following \citet{Oliver2010}. In this way, we have obtained sSFR values for each pixel in the simulated images (there are around 61,000 pixels corresponding to the galaxy) and in each of the 2.9 million dust cells as well. These values are plotted against $f_{\text{young}}$ in Fig.~\ref{fig:fyoungssfr}, for the pixels and cells of the M~81 model, and for the M~51 and M~31 pixels as reference. We observe a very similar monotonically increasing trend between $f_{\text{young}}$ and the sSFR. A power-law fit between both quantities yields the following relations:

\begin{gather}
\log f_\text{young}^\text{pix} = 0.41 \log \left(\frac{\text{sSFR}^\text{pix}}{\text{yr}^{-1}}\right) + 4.27
\\
\log f_\text{young}^\text{cells} = 0.34 \log \left(\frac{\text{sSFR}^\text{cells}}{\text{yr}^{-1}}\right) + 3.48
\end{gather}

The fit to the cell data is shown as a dotted red line in Fig.~\ref{fig:fyoungssfr}, as well as power-law fits to the M~51 and M~31 pixel data. 

A first observation from the relations above is that the slope for the M~81 pixel data exceeds the slope for the cell data. Looking at Fig.~\ref{fig:fyoungssfr}, this discrepancy seems to lie in the shape of the low-$f_\text{young}$/sSFR tail, but overall both sets of data points occupy the same space. Yet, as expected, the spread in the dust cell heating fractions is larger than the pixel heating fractions, since the range in environments is greater in the former case. What also can be noted from Fig.~\ref{fig:fyoungssfr} is that there is a clear offset between the M~81 scatter points and those found for M~51 and M~31. Some differences can certainly be expected because of the different ages for the young stellar populations used in the different studies, and because of the different methods used to estimate the sSFR. In particular, \citet{V17} used MAGPHYS pixel-by-pixel fits to quantify the sSFR, while \citet{DL14} used the MAPPINGS star-formation rates of the young ionising stellar component. This shows the needs for uniform approach to determine these quantities, in order to make future comparisons between galaxies more useful. Setting this common basis for future work is one of the goals of this present work. Despite the offset, the slope for the pixels of M~81 closely matches the slope of the M~51 and M~31 relations, with values of 0.415 and 0.481 respectively. Other offsets between cell and pixel data on one hand, and between the M~81, M~51 and M~31 points on the other hand could be resolution-related.

\begin{figure}[t]
\centering
\includegraphics[width=\columnwidth]{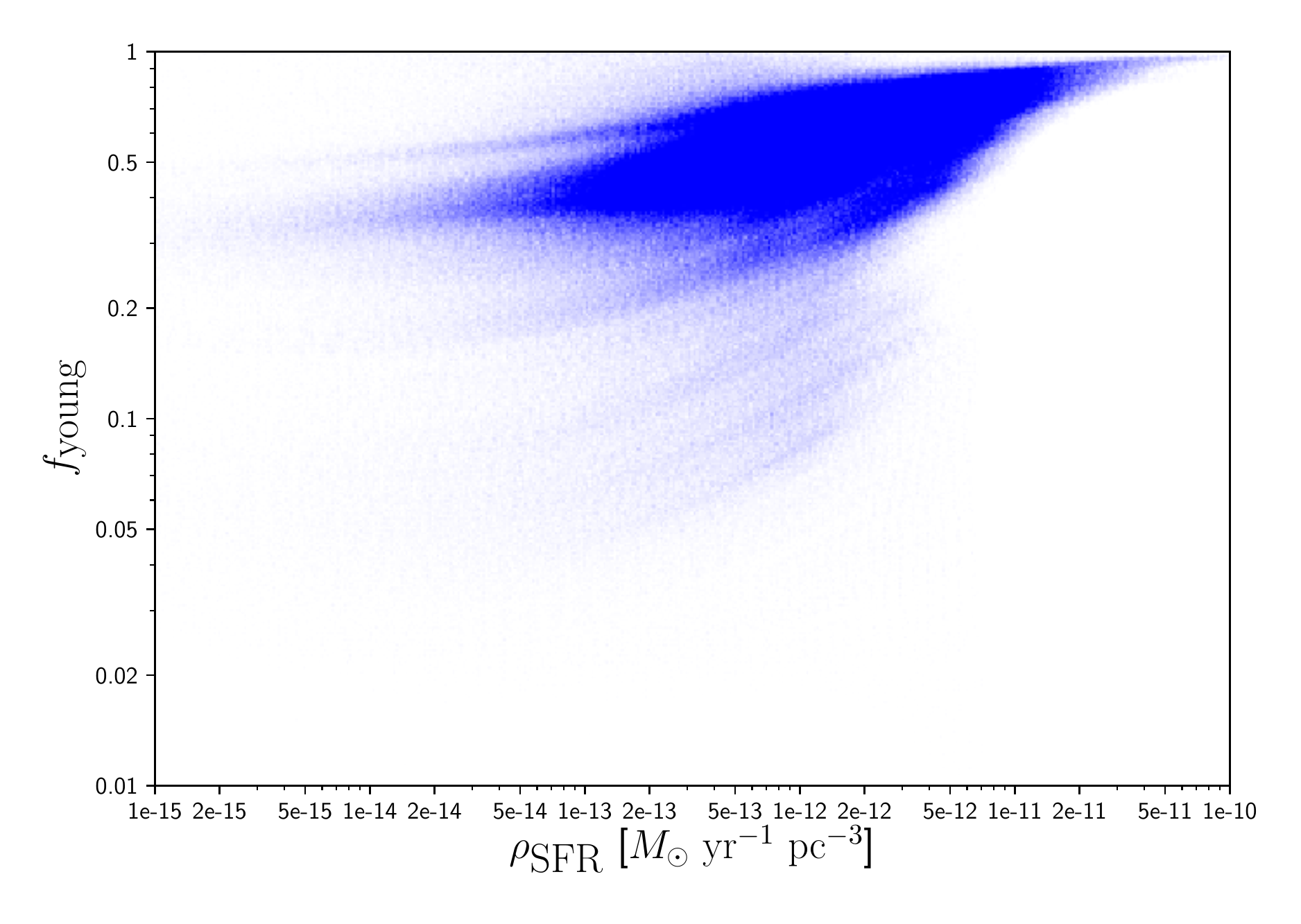}
\caption{Correlation between the star-formation rate density ($\rho_\text{SFR}$) and $f_{\text{young}}$, calculated from dust cell data of the RT model.}
\label{fig:fyoungvsfr}
\end{figure}

\begin{figure}[t]
\centering
\includegraphics[width=\columnwidth]{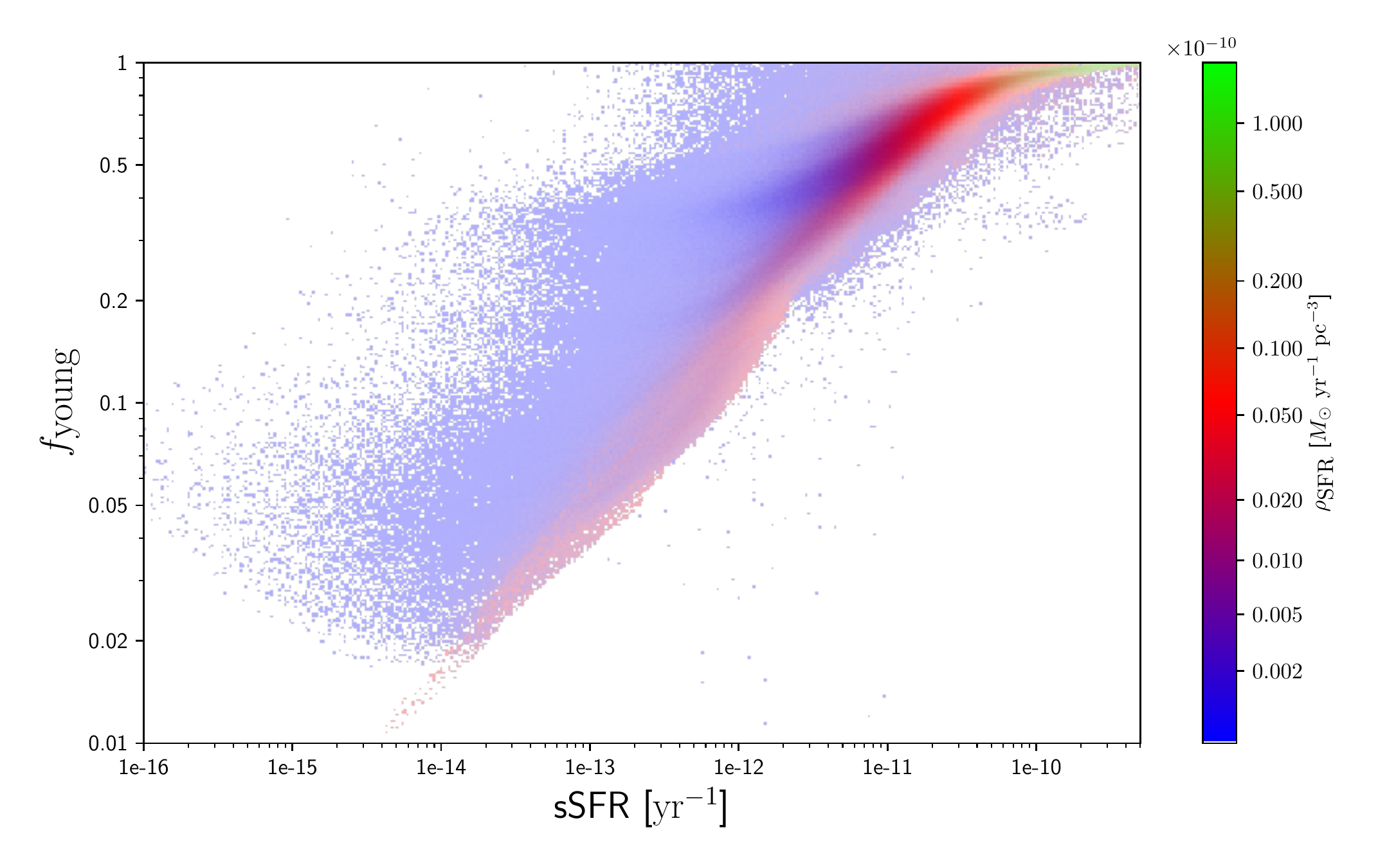}
\caption{Correlation between the specific star-formation rate and $f_{\text{young}}$, with the colour scale indicating the star-formation rate density, and the density of points represented by a transparency level.}
\label{fig:vsfraux}
\end{figure}

Our results confirm the strong relation between $f_\text{young}$ and sSFR with a Spearman rank correlation coefficient $\rho = 0.89$. We compare this with the relationship between the star formation density ($\rho_\text{SFR}$, SFR per dust cell volume) and $f_\text{young}$, plotted in Fig.~\ref{fig:fyoungvsfr}. This correlation is less clear, with a Spearman coefficient of just $\rho=0.57$. However, the star-formation rate density does seem to have an important effect on the sSFR-$f_\text{young}$ relation. In Fig.~\ref{fig:vsfraux}, we plot the same data points as in Fig.~\ref{fig:fyoungssfr}. However, in this case, we add $\rho_\text{SFR}$ as an auxiliary axis, represented by different colours. The transparency level represents the density of the scatter points. The star-formation rate density seems to explain much of the scatter in the sSFR-$f_\text{young}$ relation, with increasing $\rho_\text{SFR}$ shifting the relation towards the lower $f_\text{young}$ and higher sSFR. At a fixed sSFR, an increasing star-formation rate density decreases $f_{\text{young}}$. Equivalently, higher mass density of the old, diffuse stellar population ($=\rho_\text{SFR}/\text{sSFR}$) causes lower heating fractions, irrespective of the sSFR. We observe a very similar trend with the dust density on the auxiliary axis as compared to the star-formation rate density. This can be expected as both quantities are spatially correlated, with an enhancement in the spiral arms.

\subsection{Implications for SFR estimators}

\begin{figure*}[t!]
\centering
\includegraphics[width=0.99\textwidth]{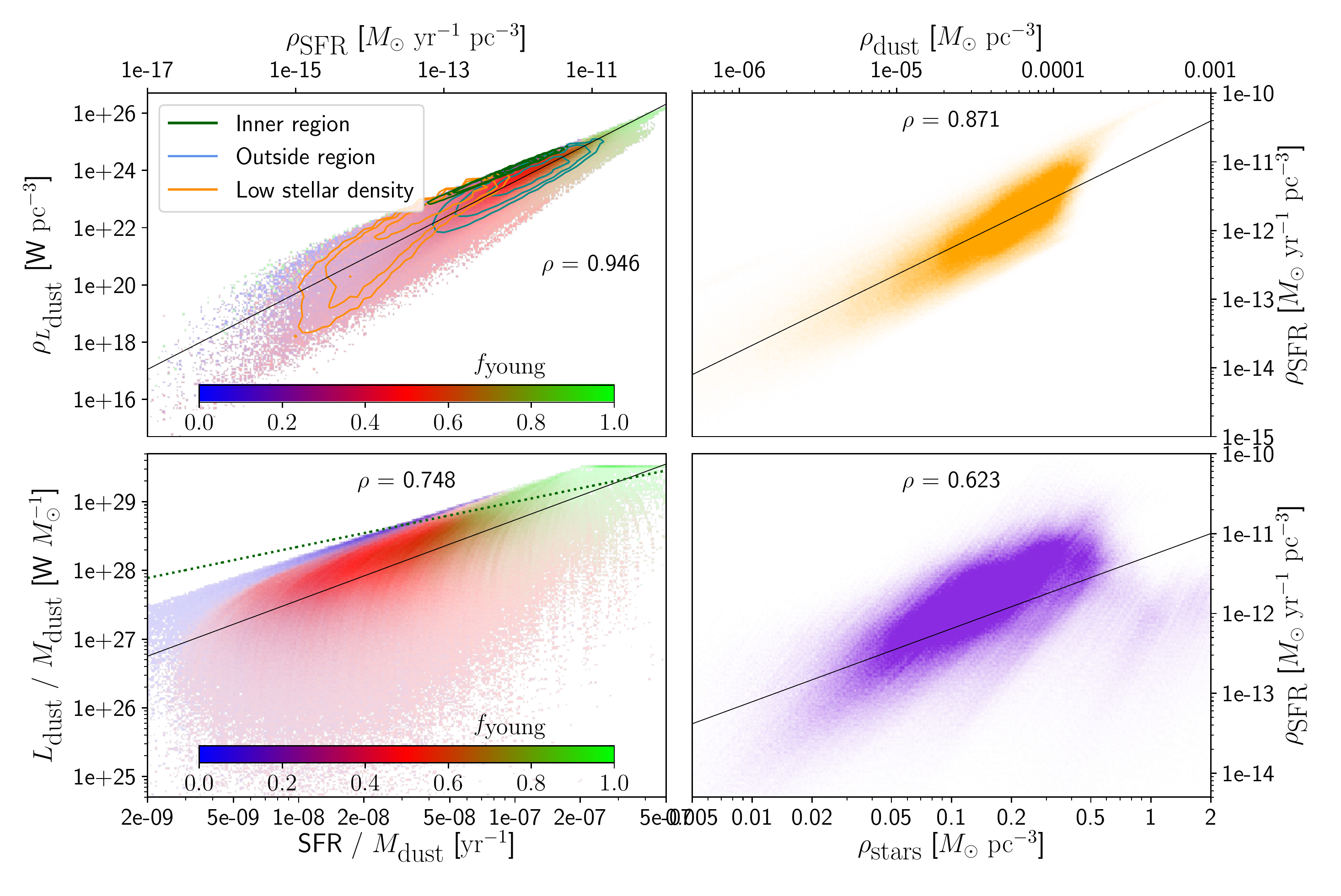}
\caption{Scatter density plots between different stellar and dust quantities from the M~81 dust cell data. Spearman rank correlation coefficients are shown at the top of each panel, and power-law fits through the data are shown as black lines. Top left panel: dust luminosity density as a function of star-formation rate density, with $f_{\text{young}}$ on the auxiliary axis. Top right panel: correlation between dust mass density and star-formation rate density. Bottom left panel: both the star-formation rate and the dust luminosity have been plotted per unit of dust mass, with the heating fraction on the auxiliary axis. The dotted green line is the power-law fit through the points with $f_\text{young} > 0.8$. Bottom right panel: correlation between the diffuse stellar mass density and the density of star-formation rate ($\rho_\text{SFR}$).}
\label{fig:dustlum}
\end{figure*}

The TIR emission is often linked to the star-formation rate in spiral galaxies and is therefore sometimes used as a direct tracer \citep{Devereux1990, Devereux1995, Buat1996, Kennicutt1998, Kennicutt2009}. The interpretation that this dust is therefore heated by young stars, however, is to be considered with caution. In the top left panel of Fig.~\ref{fig:dustlum}, we have plotted the correlation between the star-formation rate density ($\rho_\text{SFR}$) and the dust luminosity density of the dust cells in our RT model. There is a strong increasing trend between both quantities, with a Spearman correlation coefficient of 0.95. These results could lead to the conclusion that star formation is the main source of dust (TIR) emission, however we know from our analysis above that this is certainly not the case in this galaxy. 

Thanks to this analysis, we now have detailed quantitative information on the fraction of heating contributed by young stars in each dust cell, which we indicate with the colours seen in the plot. The lowest heating fractions are shown in blue, and the highest fractions are plotted in green, while intermediate heating fractions appear in red. Again, the transparency level indicates the density of points, which in this case peaks at the bright red points under the fitted power-law. The region of intermediate heating fractions (most dust cells are in this regime) coincides with the best-fitting power-law relation. Lower heating fractions also follow this relation, but with more scatter. Cells with higher heating fractions tend to lie below the relation. This means that the fitted relation could lead to underestimated star formation rates for the highest $f_\text{young}$ environments, and to highly uncertain star formation rates in low $f_\text{young}$ environments.

To illustrate how the effect of $f_\text{young}$ can be quantified on this relation based on visible or observable quantities, we indicate three different regions in the plot as contour levels. Green contour levels indicate the density of points that are in the inner (< 3~kpc) region of the model, while blue contours enclose the cells outside of this region. We have also indicated cells with low diffuse stellar density (from the 3.6~$\mu$m luminosity) with orange contours. These individual cases occupy separate but adjacent areas in the parameter space, with Spearman correlation coefficients for the respective subsets of 0.998, 0.95, and 0.88. It is evident that the fitted relation is driven by the cells in the disc where the stellar density is relatively high (i.e. the spiral arms). The inter-arm regions (orange contours) match the scatter in the low $f_\text{young}$ regime. Bulge cells are significantly offset from the relation.

Even including the considerable scatter in the dust luminosity to star formation relation, driven by $f_{\text{young}}$, the correlation is still strong. However, as noted by other authors, the luminosity of a dust distribution is determined by both its density and its heating rate (or equivalently, temperature). Regions of higher dust mass absorb more stellar photons, and thus also emit more energy in the same volume. It is presumed that the dust density is tightly linked to the density of gas, and star formation is linked to the gas content via the Schmidt law \citep{Schmidt1959, Kennicutt1998}, the density of dust is indirectly correlated to the star formation rate density \citep{Bendo2012}. Indeed, this effect is clear from the top right panel in Fig.~\ref{fig:dustlum}, where we have plotted the correlation between the dust density and the density of star-formation rate ($\rho_\text{SFR}$). There is a strong relation between both quantities in our 3D model, with a correlation coefficient of 0.87, which clearly has a considerable effect on the relation shown in the first panel. 

To eliminate the influence of the dust density, and thus also its link to the star-formation rate density, we make a different version of the first plot on the bottom left panel of Fig.~\ref{fig:dustlum}. Here, we divide the dust luminosities in each cell by the corresponding dust masses, instead of the cell volumes. To make this quantity comparable to the star formation, we also normalise these star-formation rates to the dust masses. The resulting correlation is weaker than the dust luminosity density versus $\rho_\text{SFR}$ comparison above, with a Spearman correlation coefficient of 0.75. If star formation were the main driver of dust heating, and this heating is purely local, one would expect a strong relation between both quantities. We therefore show $f_\text{young}$ on the auxiliary axis, and fit the highest $f_\text{young}$ (> 0.8) set of points with a power-law (dotted green line). The Spearman correlation coefficient for these points is 0.79. This is only slightly higher than the correlation for all of the data points, which makes us postulate that non-local heating has a strong effect. In general, we argue that the apparent relation between star formation and TIR luminosity must be interpreted with caution, especially on local scales.

\subsection{Dust temperature}

With dust luminosities increasing with star formation rate density (see top left panel of Fig.~\ref{fig:dustlum}), one could expect to find the highest dust temperatures in the high SFR regions. However, much of the elevated dust luminosity can be traced to an increased dust density, as we have shown in the bottom left panel of Fig.~\ref{fig:dustlum} where we have plotted the dust luminosity per unit of dust mass as a measure of the internal energy density of the dust. Even though this scatter plot shows only a weak trend which can also partially be explained by increasing diffuse stellar densities, one could still expect the star formation causing the highest dust temperatures in the galaxy. 

\begin{figure}[t!]
\centering
\includegraphics[width=\columnwidth]{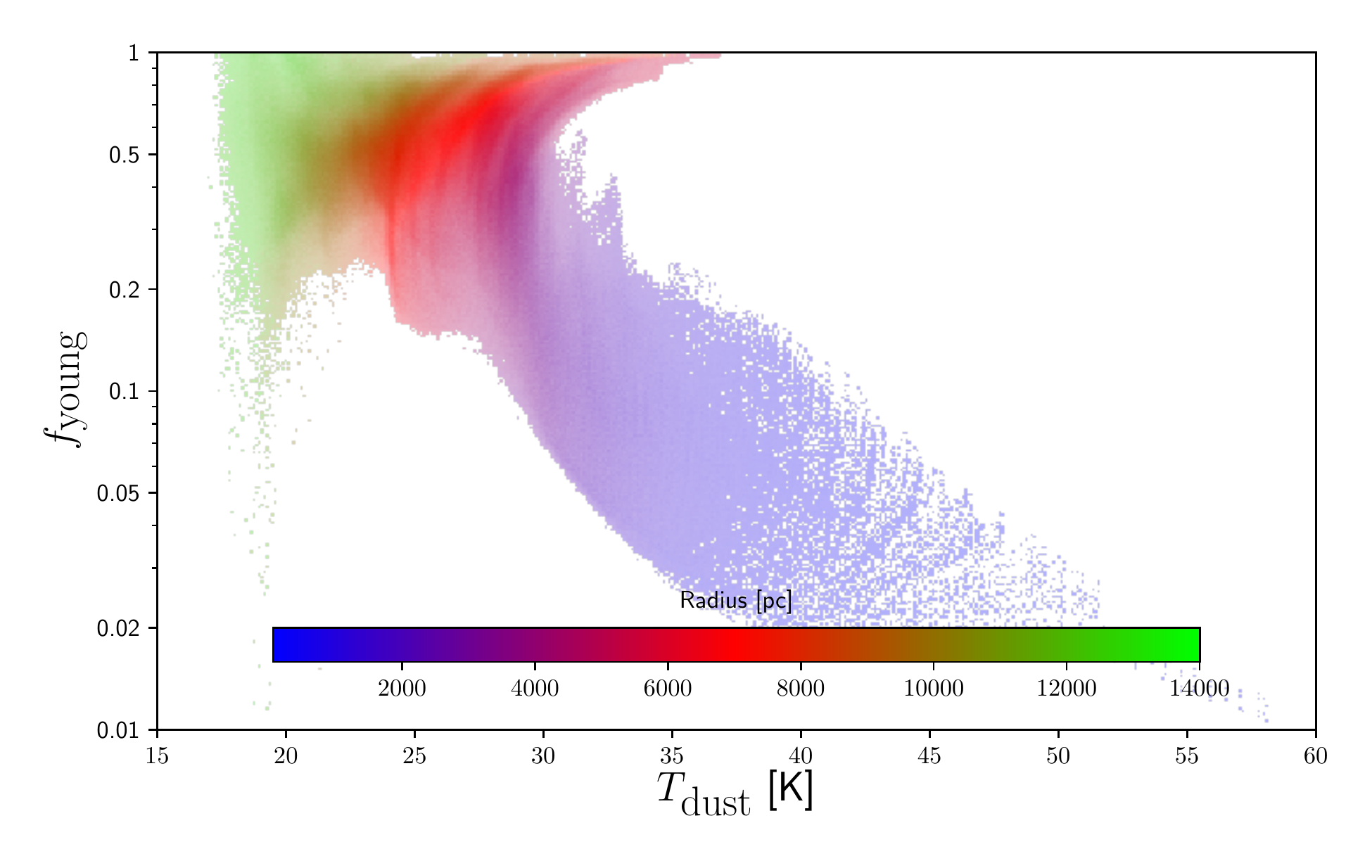}
\caption{Correlation between the diffuse dust temperature and $f_{\text{young}}$. The galactocentric radius in the plane of the galaxy has been plotted on the auxiliary colour axis.}
\label{fig:temperature}
\end{figure}

In Fig.~\ref{fig:temperature}, we plot $f_{\text{young}}$ as a function of the dust temperature, showing that - counterintuitively - the highest heating fractions correspond to the lowest dust temperatures and vice versa. There is in fact an anti-correlation between the two quantities, with a Spearman rank coefficient of $-0.095$ (the high temperature tail is sparse in points). We must stress, however, that the dust temperatures used here are those of the diffuse dust, and including the temperature of the dense dust clouds in the star-formation regions would add a certain set of high SFR, high temperature points. These temperatures are not easily included as they are subgrid properties of the MAPPINGS template. Still, with the radius added on an auxiliary axis as we have done in Fig.~\ref{fig:temperature}, it is meaningful to see how much the central old stellar bulge is heating the diffuse dust. None of the diffuse dust in the disc of the galaxy is heated to such extent. Even though the diffuse dust is shielded from the hardest stellar radiation by the subgrid clumps of dust, one could expect the diffuse dust to still be very effectively heated in the spiral arms of the galaxy. This analysis, however, confirms that the old stars in M~81 not only heat the central 3~kpc region of the galaxy almost exclusively as compared to young stars, but also significantly in absolute numbers.

\subsection{Limitations and caveats in our modelling approach}

The results we have described and discussed in this section, are all based on the modelling methodology presented in Sec.~{\ref{sec:approach}}. In the ideal world, this modelling would be a perfect radiative transfer inversion that translates the observed, dust-affected images to intrinsic 3D distributions of stars and dust. In order to make the problem tractable, we are forced to make a number of simplifying assumptions and choices. On the one hand, these assumptions drastically reduce the number of free parameters to be determined in the inversion problem. Instead of a completely intractable inversion problem, they lead to an optimisation problem with only three free parameters to be determined (Sec.~{\ref{ssec:optimisation}}). While this seems reasonably simple, it remains a challenging problem, given the computational cost of a single panchromatic radiative transfer simulation. On the other hand, these assumptions are simplifications that do not perfectly represent reality, and they induce potential systematic errors in our method.

One such simplification is to constrain the spatial distributions of the different stellar populations to fixed spatial templates, each with a fixed spectral emission template. We have chosen to use only four different stellar components, and to characterise the intrinsic spectrum of each of these by an SSP model corresponding to a single age and a single metallicity. The derivation of the spatial distribution of each of these components is based on empirical recipes that take the observed surface brightness distributions  in different bands as input. The majority of stars in galaxies form in star clusters, and these clusters eventually disperse and the individual stars diffuse into the surrounding environment on time scales of the order to tens to hundreds of Myrs, becoming part of the larger galactic background. Empirically, this can be seen from detailed spatially resolved star formation history reconstructions in nearby galaxies \citep[e.g.,][]{1999AJ....117.2831H, 2008MNRAS.391L..93G, 2009MNRAS.392..868B, 2011MNRAS.412.1539B, 2015ApJ...805..183L}. This shows that our representation of the young stellar population by just a superposition of two components with ages of 10 and 100 Myr is clearly a simplification. Similarly, we have assumed a single geometric distribution for the old stellar disc, with a fixed stellar age of 8~Gyr. However, according to the current $\Lambda$CDM galaxy disc formation scenarios \citep{1991ApJ...379...52W, 1998MNRAS.295..319M, 2006ApJ...645..986R}, one might expect large-scale, i.e., kpc-scale, intrinsic stellar population gradients in galactic discs. Indications for such intrinsic gradients have been obtained empirically from colour gradients \citep{1996A&A...313..377D, 2004ApJS..152..175M, 2007ApJ...658.1006M, 2016AJ....151....4D} and gradients in color-magnitude diagrams of nearby disc galaxies \citep{2009ApJ...695L..15W, 2010ApJ...712..858G}. The neglect of these gradients in our modelling might also affect our modelling results.

Another aspect in our modelling that could potentially lead to significant systematic errors relates to the interstellar dust component. Firstly, we have assumed that the interstellar dust has uniform optical properties throughout the galaxy, whereas there are now various indications that the dust properties in galaxies are not uniform. For example, spatially resolved studies of the infrared emission in M~31 \citep{2012ApJ...756...40S, 2014ApJ...780..172D, 2019MNRAS.489.5436W} and M~33 \citep{2014A&A...561A..95T, 2018A&A...613A..43R} strongly suggest a gradient in the physical properties of the dust grains. Recent spatially resolved studies of the dust properties in nearby DustPedia galaxies corroborate this claim \citep{2019A&A...623A...5D, 2019MNRAS.489.5256C}. Secondly, we have assumed that the spatial distribution of dust can adequately be derived from the FUV attenuation map, which we derived using the calibration obtained by \citet{Cortese2008}. Compared to a dust mass distribution based on infrared SED fits, as used for the radiative transfer modelling of M~31 \citep{V17} and M~33 \citep{2019MNRAS.487.2753W}, this approach has the advantage that a higher spatial resolution can be obtained \citep[see discussion in][]{DL14}. On the other hand, this approach has some disadvantages as well. The \citet{Cortese2008} calibration of the UV attenuation is based on a simple radiative transfer model solution for a sandwich model by \citet{2003A&A...402...37B}. This is obviously a simplification of the real situation, and it is not immediately clear whether this approximation is acceptable on local scales in all environments, and how this systematically affects the results of the modelling.

Finally, our modelling recipe contains a number of other ingredients and assumptions that might systematically affect the results. We have assumed that each of the different components, apart from the stellar bulge, has a perfectly exponential behaviour in the vertical direction, with scale height ratios determined by observations in nearby galaxies. Furthermore, we have assumed that the intrinsic luminosity of bulge and old stellar disc can directly be obtained from the IRAC 3.6~$\mu$m emission, which implicitly assumes that the emission is unaffected by dust attenuation or emission, and that the young stellar populations have a negligible contribution to the IRAC 3.6~$\mu$m band. These conditions may not always be satisfied for all galaxies \citep[e.g.,][]{Camps2018}.

There are a few indications where we see the above limitations play up (see Sect.~\ref{SEDComparison.sec} and Sect.~\ref{ImageComparison.sec}). The fact that our model tends to underestimate the FIR flux in the spiral arms may lead to an underestimation of the heating by young stars. On the other hand, the 24~$\mu$m flux of the model is too high, which can work in the opposite direction. At the same time this discrepancy challenges the chosen parameter set for the ionising stellar template. Furthermore, the relatively low $g$ band flux may cause an underestimation of heating by older stellar populations. It could even point towards heating by stellar populations of intermediate age.

The bottomline is that the simplifying assumptions we had to make also induce systematic errors on the modelling results. Unfortunately, such systematic errors are very hard to quantify exactly. One option to quantify the systematic errors that result from some assumptions would be to gradually adapt/refine our modelling and check the effects of this gradual refinement on the results. For example, we could, in principle, gradually increase the number of stellar components or introduce a radial stellar population gradient or dust properties gradient in the model, and check how this affects the modelling results. Each additional parameter in the modelling significantly increases the computational cost of the modelling, however, which implies that our room for manoeuvre is limited. Moreover, it can be expected that several of the simplifying assumptions are coupled. For example, changes to the way the FUV attenuation map is calculated will affect the dust distribution, but also the distribution of the young stellar populations. The predictive value of ``small'' independent tests could therefore possibly be limited. 

We believe that the most promising way to map the systematic errors induced by the simplifying assumptions in our modelling approach is to apply the technique to simulated galaxies, and compare the model results to the intrinsic results obtained from the simulation itself. The advantage of this approach is that all of the explicit and implicit assumptions made in our method can, in principle, be investigated adequately and simultaneously. The necessary condition is that realistic galaxy models be available as a starting point for this investigation. Up to relatively recently, such models were non-existent: simulations of the formation of late-type spiral galaxies in a $\Lambda$CDM framework have traditionally failed to yield realistic candidates. Only in recent years, hydrodynamical simulations have started to successfully reproduce late-type spiral galaxies with realistic sizes, angular momentum, thicknesses and scaling properties \citep[e.g.,][]{2011ApJ...742...76G, 2014MNRAS.437.1750M}. In particular, current state-of-the-art cosmological high-resolution zoom simulations as FIRE-2 \citep{2018MNRAS.480..800H, 2018MNRAS.481.4133G}, Auriga \citep{2017MNRAS.467..179G}, and NIHAO \citep{2015MNRAS.454...83W, 2019arXiv190905864B} produce simulated disc galaxies that closely reproduce observed disc galaxies in many aspects. 

As an important future work, we will apply our inverse radiative transfer modelling approach to a set of high-resolution simulated disc galaxies from the Auriga simulation. We will first generate mock observations for these simulated galaxies, using the SKIRT-based pipeline developed in the frame of the EAGLE simulation \citep{Camps2016, Camps2018, Trayford2017}. This pipeline contains a recipe to insert interstellar dust in simulated galaxies based on a fixed dust-to-metal ratio, a resampling technique to increase the time and spatial resolution of recent star formation, and a radiative transfer treatment to generate mock UV-submm broadband images from arbitrary viewing points. Subsequently we will apply the methodology outlined here to these mock data, and compare the modelling results directly to the intrinsic simulation properties. We are convinced that this approach will quantify the importance of the systematic errors that result from the simplifications and choices in our modelling, and if necessary, will guide us towards improvements in our approach. In the meantime, we urge to interpret our results with the necessary caution, and keep the caveats in our methodology into account.

\section{Conclusions}
\label{sec:conclusions}

Based on morphological constraints from observed image data, assumptions about the composition and vertical distributions of stars and dust, we have created 3D radiative transfer models of the early-type spiral galaxy M~81. These models have been evaluated by matching their simulated imagery to observations across the entire UV to submm wavelength range, yielding a best fitting description of the galaxy. We have set guidelines for the consistent creation of such RT models, which can be applied to other well-resolved dusty galaxies. The different steps of our method are publicly available as python modules. The model constructed for M~81 fits the observed data well as proven by residuals which are mostly within 50\%. Based on a 3D study of the absorption and re-emission of stellar radiation by dust in the M~81 model, we reach the following conclusions:\\

\begin{itemize}
  \item Star formation is not necessarily the main driver of dust heating in galaxies, as only 50.2\% of the absorbed energy is attributed to young stellar populations in our M~81 model.
  
  \item In total, about 11.3\% of the stellar radiation is absorbed (and re-emitted) by dust, but this number is 62\% for the radiation from the young ionising stellar population and 30\% for the young non-ionising population. Old stars have around 7\% of their energy absorbed. In the FUV band, 53\% of the stellar energy is reprocessed by dust ($A_\text{FUV}$ = 0.69~mag).
  
  \item The fraction of heating attributed to young stars varies significantly within the galaxy, with values of < 1\% in the center to about 60\% in the spiral arms. These values are invariant to the height above the midplane of the galaxy, to within a few percent.
  
  \item The dust heating in the central region of M~81 is dominated by the old stellar population. The fraction of absorbed energy by young stellar populations in a particular wavelength drastically drops between the UV wavebands and the SDSS~\textit{g} band. This is reflected throughout the entire disc of the galaxy, but local increases of the heating fraction by young stars at the spiral arms are visible.
  
  \item A strong correlation between the heating fraction by young stars, $f_{\text{young}}$, and the specific star-formation rate (sSFR) points to this sSFR as the favoured diagnostic in determining the relative contribution of old and young stellar populations to the absorption by dust. This correlation gives an observational tool to probe the dust heating fraction, in cases where a full RT model can be obtained or if this is not the intent. The benefit of having the sSFR as a probe to the intrinsic properties of the radiation field is that it has a close connection to observational UV-optical colours.
  
  \item The star-formation rate density causes much of the scatter to the power-law relation between $f_\text{young}$ and sSFR, meaning that the density of old stellar stars has a secondary effect to the correlation. There is a strong correlation between the dust luminosity density and the star-formation density, but this relation is mainly driven by the tight link between dust density and star formation density. Regions with different heating fractions are offset on the dust luminosity density -- star-formation density plane.
  
  \item When plotting the dust luminosity and star formation rate per unit of dust mass, the correlation is much weaker. We show that part of the remaining trend could be caused by a correlation between old stellar density and star formation rate density. On average, higher old stellar densities in the disc are found with slightly higher star formation, and both contributing about half to the dust heating causes the described trend with dust luminosity. 
  
  \item There is almost no correlation between the diffuse dust temperature and the fraction of heating attributed to young stars. The dominant bulge in M~81 accounts for the highest dust temperatures in the model, and thus these high temperatures occur in the region where the heating fraction by young stellar populations is lowest. This fraction is low in the central 3~kpc region in all of the absorption bands looked at.
\end{itemize}

M~81 is the first galaxy that was modelled following to the instructions detailed in this paper. The semi-automated and modular approach can be readily adopted to model other galaxies, with the caution that differences in available data should allow flexibility in the pipeline. We also plan to apply our modelling technique to mock data corresponding to a set of hydrodynamically simulated  disc galaxies from the Auriga simulations to assess the importance of systematic errors in our modelling approach. Finally, we encourage interested readers to further explore our modelling approach. The necessary configuration files to run the model in SKIRT are available on the DustPedia archive (see Appendix~{\ref{AppendixC.sec}} for details).

\bibliographystyle{aa}
\bibliography{AA-2019-35770}

\section*{Acknowledgements}

We thank the anonymous referee for his/her very extensive and detailed referee report, which significantly improved the presentation in this paper.

DustPedia is a project funded by the EU (grant agreement number 606847) under the heading "Exploitation of space science and exploration data (FP7-SPACE-2013-1)" and is a collaboration of six European institutes: Cardiff University (UK), National Observatory of Athens (Greece), Instituto Nazionale di Astrofisica (Italy), Universiteit Gent (Belgium), CEA (France), and Universit\'e Paris Sud (France). 

MB, IDL and AT gratefully acknowledges the support of the Flemish Fund for Scientific Research (FWO-Vlaanderen). SV acknowledges the support of the UGent-BOF and the Flemish-FWO research funds. AVM expresses gratitude for the grant of the Russian Foundation for Basic Researches number mol\textunderscore a 18-32-00194. 

This research made use of Astropy,\footnote{http://www.astropy.org} a community-developed core Python package for Astronomy \citep{AstropyI, Astropy2018}. This research has made use of the NASA/IPAC Infrared Science Archive (IRSA; \url{http://irsa.ipac.caltech.edu/}), and the NASA/IPAC Extragalactic Database (NED; \url{https://ned.ipac.caltech.edu/}), both of which are operated by the Jet Propulsion Laboratory, California Institute of Technology, under contract with the National Aeronautics and Space Administration.

The computational resources (Stevin Supercomputer Infrastructure) and services used in this work were provided by the VSC (Flemish Supercomputer Center), funded by Ghent University, FWO and the Flemish Government -- department EWI.

\appendix

\section{Custom aperture photometry}
\label{sec:appendix_photometry}

Table~\ref{tab:fluxes} summarises the properties of the image data used for the radiative transfer modelling, as well as the final aperture photometry fluxes extracted from them. 

\begin{table*}
\caption{Summary of the properties of the image data and photometry results for this work}
\label{tab:fluxes}
\centering
\begin{tabular}{llccccc}
\hline\hline \\[-1mm]
band & $\lambda_{\text{eff}}$ & pixel scale & FWHM & flux density \\
 & [$\mu$m] & [arcsec] & [arcsec] & [Jy] \\[2mm] \hline \\[-1mm]
{\textit{GALEX}} FUV & 0.154 & 3.2 & 4.48 & 0.128 $\pm$ 0.006 \\
{\textit{GALEX}} NUV & 0.227 & 3.2 & 5.05 & 0.184 $\pm$ 0.005 \\
SDSS $u$ & 0.359 & 0.45 & 1.3 & 0.963 $\pm$ 0.0125 \\
SDSS $g$ & 0.464 & 0.45 & 1.3 & 4.16 $\pm$ 0.03 \\
SDSS $r$ & 0.612 & 0.45 & 1.3 & 8.46 $\pm$ 0.07 \\
SDSS $i$ & 0.744 & 0.45 & 1.3 & 12.53 $\pm$ 0.09 \\
SDSS $z$ & 0.890 & 0.45 & 1.3 & 15.03 $\pm$ 0.12 \\
2MASS J & 1.235 & 1.0 & 2.0 & 19.19 $\pm$ 0.54 \\
2MASS H & 1.662 & 1.0 & 2.0 & 23.25 $\pm$ 0.65 \\
2MASS Ks & 2.159 & 1.0 & 2.0 & 17.91 $\pm$ 0.50 \\
{\textit{WISE}} W1 & 3.352 & 1.37 & 5.79 & 10.0 $\pm$ 0.3 \\
IRAC I1 & 3.508 & 0.75 & 1.9 & 10.85 $\pm$ 0.33 \\
IRAC I2 & 4.437 & 0.75 & 1.81 & 6.5 $\pm$ 0.2 \\
{\textit{WISE}} W2 & 4.603 & 1.37 & 6.37 & 5.9 $\pm$ 0.2 \\
IRAC I3 & 5.628 & 0.6 & 2.11 & 6.23 $\pm$ 0.19 \\
IRAC I4 & 7.589 & 0.6 & 2.82 & 8.47 $\pm$ 0.25 \\
{\textit{WISE}} W3 & 11.56 & 1.37 & 6.6 & 5.6 $\pm$ 0.3 \\
{\textit{WISE}} W4 & 22.09 & 1.37 & 11.89 & 5.52 $\pm$ 0.31 \\
MIPS 24 & 23.21 & 1.5 & 6.43 & 5.05 $\pm$ 0.25 \\
MIPS 70 & 68.44 & 4.5 & 18.74 & 76.45 $\pm$ 7.64 \\
PACS 70 & 68.92 & 2.0 & 5.67 & 77.26 $\pm$ 5.41 \\
MIPS 160 & 152.6 & 9.0 & 38.78 & 305.89 $\pm$ 36.71 \\
PACS 160 & 153.9 & 4.0 & 11.18 & 262.35 $\pm$ 18.36 \\
SPIRE PSW & 247.1 & 6.0 & 18.15 & 140.02 $\pm$ 7.70 \\
SPIRE PMW & 346.7 & 8.0 & 24.88 & 69.39 $\pm$ 3.82 \\
HFI 857 & 349.9  & 102.1 & 278.0 & 59.82 $\pm$ 3.83 \\
SPIRE PLW & 496.1 & 12.0 & 36.09 & 30.98 $\pm$ 1.70 \\
HFI 545 & 550.1  & 102.1 & 290.0 & 19.62 $\pm$ 1.20 \\
HFI 353 & 849.3 & 102.1 & 296.0 & 5.25 $\pm$ 0.04 \\
HFI 217 & 1382 & 102.1 & 301.0 & 0.584 $\pm$ 0.001 \\
HFI 143 & 2096 & 102.1 & 438.0 & 1.040 $\pm$ 0.001 \\
HFI 100 & 2998  & 205.7 & 581.0 & 0.455 $\pm$ 0.001 \\
LFI 070 & 4283 & 205.7 & 799.0 & 0.324 $\pm$ 0.001 \\
LFI 044 & 6813 & 205.7 & 1630.0 & 0.114 $\pm$ 0.001 \\
LFI 030 & 9993 & 205.7 & 1940.0 & 0.0546 $\pm$ 0.0002 \\[3mm]
\hline
\end{tabular}
\end{table*}

\section{Wavelength grid}
\label{sec:appendix_wavelengthgrid}

\begin{figure*}[t!]
\centering
\includegraphics[width=\textwidth]{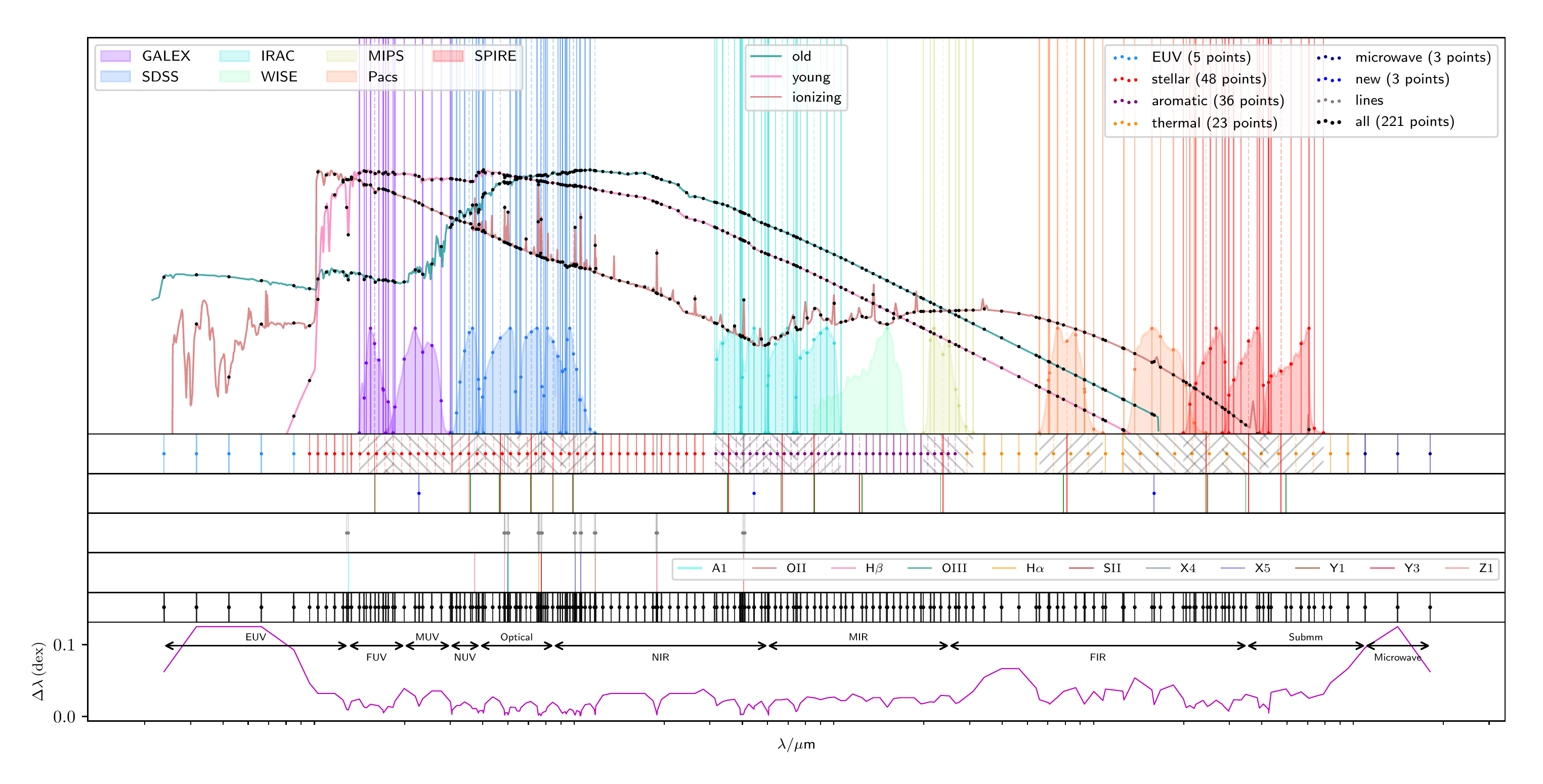}
\caption{Wavelength grid used for the RT simulations. The top panel shows the 3 different stellar template spectra used for the models, on an arbitrary scale and normalised independently to match the frame. Rendered on top of these curves are black dots that show the points of the wavelength grid, to illustrate how they sample the intrinsic spectra of our model components. Also shown are the transmission curves of the filters for which we have observed data and for which we create mock observations. The coloured vertical lines plotted on top of these response curves show the additional sampling that is performed for each band, also shown as dots. The second panel shows the sampling of the different regimes with different densities, in a different colour per regime. The grid points with dashed vertical lines, and crossed out with the diagonal pattern, are removed and replaced by the wavelengths added for the sampling of the above filter response curve, because the number of points within the filter range was not sufficient. The next horizontal panel shows the wavelength grid points that have been placed at the effective wavelengths of the different bands. When this wavelength is close to an existing wavelength point, it replaces the original one (red and green lines), otherwise it is added as an additional wavelength (blue line and dot). The fourth panel shows the extra sampling that has been performed for prominent absorption and emission line features in the stellar spectra, adding a grid point at the peak (or valley) of the feature and at its edges. The position and origin of the corresponding line features are indicated in the panel below. The resulting wavelength grid is shown in the 6th panel. The density of the total grid in logarithmic space is plotted in the bottom panel, showing that the distance between any subsequent wavelengths in the grid is well below 0.05~dex, except in the EUV where the stellar emission is insignificant.}
\label{fig:wavelengths}
\end{figure*}

The spectral discretisation of the Monte Carlo RT simulations is realised by evaluating all interactions on photon packages or luminosity packages within certain (preferably small) wavelength bins. These luminosities are converted into spectral densities again for the simulation output data cubes, based on the central wavelengths of the bins. For panchromatic radiative transfer modelling purposes, logarithmic and nested logarithmic have predominantly been used \citep{DL14, V17}, both of which are implemented in SKIRT. For our purposes however, we prefer to perform more fine-tuning on the precise placement of the wavelength grid points, for several reasons:

\begin{enumerate}
\item We want a good sampling of the different input stellar spectra and expected dust emission features, while avoiding placing too many grid points in the smooth parts of these spectra.

\item Similarly, we also want to give differential weight to parts of the wavelength range that are less or more important to the energy balance of a galaxy.

\item Because we perform spectral convolution on the output simulated SEDs, we take care of the sampling of each individual filter's wavelength range in the grid to assure realistic results.

\item Lastly, to precisely study the energy budget of the different stellar components, we want to sample the sharp emission and absorption line features present in the stellar templates well, because misplaced wavelength grid points may erase or broaden such features.
\end{enumerate}

The construction of the wavelength grid used for our RT calculations thus goes as follows. We define five wavelength regimes within the galaxy spectrum, that each have a distinct, recognisable footprint that can be originated to a specific component and/or physical process. We define the extreme UV (EUV) regime (emission solely attributed to hot, young stars) from 0.02 to 0.085~$\mu$m, the stellar regime (containing the bulk of the diffuse stellar emission) from 0.085 to 3 $\mu$m, the aromatic regime (containing the emission features from very small grains (VSGs)) from 3 to 27~$\mu$m, the thermal dust regime from 27 to 1000 $\mu$m, and the microwave regime (cold dust) from 1000 to 2000~$\mu$m. We assign a weight of only 8\% to the EUV (because the emission is significantly lower in this regime), 30\% to the stellar part, 38\% to the aromatic regime (we give a high weight to this part of the spectrum to irregularity and thus limit numerical digressions in the RT calculations), about 15\% for the thermal dust part (which is smooth and well-behaved), and only 8\% to the microwave spectrum (since we are not really interested in this part of the spectrum nor does it influence the preciseness of the RT simulation). The weight defines the density of the grid at a particular regime; i.e. the number of points per decade. We distribute a total of 115 wavelength points across the entire spectrum, by creating a logarithmic range within each regime. The sampling in each of these regimes is shown in Fig.~{\ref{fig:wavelengths}} in the second horizontal panel. We check whether each observed filter's wavelength range is sampled sufficiently by imposing a minimal number of points within. If necessary, we render a new logarithmic subgrid with the acceptable number of points within the waveband range and replace the original points in that range. This is also well illustrated in Fig.~{\ref{fig:wavelengths}}. We accommodate for the most prominent emission and absorption line features in the template spectra by adding a wavelength point just before, at the peak or trough, and just after the feature. Fig.~{\ref{fig:wavelengths}} also shows the density of the resulting wavelength grid (bottom panel), which contains about 250 points.

\section{Reproducing the results}
\label{AppendixC.sec}

In the spirit of open and reproducible science, we publicly release the SKIRT input file ({\tt{ski}} file), together with all of the input data, for our M~81 simulation. As SKIRT itself is also publicly available, any user can rerun the simulation and reproduce or double-check our results. Moreover, users are welcome to modify the {\tt{ski}} file to extract that kind of information that is relevant to his/her scientific interest, or to apply it to other galaxies. All information can be found on the DustPedia archive, under the heading ``SKIRT''.

\section{global vs local $\chi^2$}
\label{app:localglobal}

In our methodology, we have used a combination of optimisation metrics (see Sect.~\ref{ssec:optimisation}). To evaluate the parameter combinations of the first exploratory grid, a global $\chi^2$ value was computed using integrated SED fluxes. The second iteration in the optimisation relied on a local, pixel-by-pixel SED comparison to construct the $\chi^2$. The advantages of the latter metric are obvious: resolved differences in the images can be more accurately reflected in the goodness-of-fit. There's a risk that these (partly) cancel out when using a global $\chi^2$. The main disadvantage of computing the local $\chi^2$ is the need of model images that are convolved to the same spectral and spatial resolution as the data, and are of sufficiently high signal-to-noise. This requires significantly more computing time: a factor of $\sim 10$ in the case of M81 using the latest version (v8) of SKIRT.

We expect that the computational barriers will become less prohibitive in the future. In particular, SKIRT 9 (Camps \& Baes, submitted) allows for direct generation of broadband images, without the expensive intermediate step of generating images across a narrow-bin wavelength grid. Still, it is possible that future users want to use a global $\chi^2$ for their particular optimisation. For example, when a large set of galaxies requires modelling, or when the data don't allow the computation of a local $\chi^2$. To test this situation, we also computed the global $\chi^2$ for the 343 models in the second optimisation step. This yielded a different best-fit model, which we ran through the same dust heating analysis pipeline.

Table~\ref{tab:comparison} compares the outcome of some key properties for both models and lists their relative difference. There are appreciable differences between the three free parameters of both best-fit models. The total dust mass is rather similar with only a 0.06 dex difference. The young stellar populations are significantly more offset, but in opposite direction. While the non-ionising component has a luminosity that is 0.25 dex lower for the local $\chi^2$ model, the ionising component is 0.75 dex higher. Still, in comparison with the luminosity of the old stellar component, these are relatively moderate changes. It does highlight a degeneracy between the both components of the young stellar populations when using only the global $\chi^2$ in the optimisation. As a result, the effects on the dust heating properties are very small. In fact we retrieve the exact same $f_\text{abs}$ of $11.3 \%$. The dust heating fraction is only marginally lower (3.8 percentage points) for the model with the global $\chi^2$.

This experiment suggests that the use of a global $\chi^2$ increases the risk of degeneracies between the free parameters in the model. In the case of M81 this can be due to a relatively strong old stellar population, which in any case hampers the constraints on the young stellar populations. Fortunately, the dust heating properties remain unchanged as do the conclusions drawn from this work. Still, it is not certain that this will be the case for other galaxies. The better way is to compute the local $\chi^2$ during the optimisation process.

\begin{table}
\caption{Comparison of $f_\text{abs}$, mean global $f_\text{young}$, and best fit values.}
\begin{center}
\scalebox{0.9}{
\begin{tabular}{lcc|c}
\hline 
\hline 
Quantity              & $\chi^2_\text{global}$ & $\chi^2_\text{local}$ & diff. [dex] \\
\hline
$f_\text{abs}$~[\%]                                           & 11.3  & 11.3 &   0.00 \\
$f_\text{young}$~[\%]                                         & 48.4  & 50.2 &  -0.02 \\
$M_\mathrm{dust}~[10^7~M_\odot]$                              & 1.48  & 1.28 &   0.06 \\
$L_\mathrm{yni,~FUV}~[10^{36}~{\text{W}}~\mu{\text{m}}^{-1}]$ & 5.66  & 3.18 &   0.25 \\
$L_\mathrm{yi,~FUV}~[10^{36}~{\text{W}}~\mu{\text{m}}^{-1}]$  & 0.28  & 1.59 &  -0.75 \\
\hline \hline
\end{tabular}}
\label{tab:comparison}
\end{center}
\end{table}

\end{document}